\documentclass[entropy,review,submit,moreauthors,pdftex]{Definitions/mdpi} 

\firstpage{1}
\pubvolume{xx}
\issuenum{1}
\articlenumber{5}
\pubyear{2020}
\copyrightyear{2020}
\history{Received: date; Accepted: date; Published: date}

\preto{\abstractkeywords}{\nolinenumbers}

\usepackage{amssymb}

\Title{Dynamics of ion channels via non-Hermitian quantum mechanics}

\Author{Tobias Gulden $^{1}$\orcidA{} and Alex Kamenev $^{2,3,}$*\orcidB{}}

\AuthorNames{Tobias Gulden and Alex Kamenev}

\address{%
	$^{1}$ \quad IST Austria, 3400 Klosterneuburg, Austria; tgulden@ist.ac.at\\
	$^{2}$ \quad School of Physics and Astronomy, University of Minnesota, Minneapolis, Minnesota 55455, USA; kamenev@physics.umn.edu
	$^{3}$ \quad William I. Fine Theoretical Physics Institute, University of Minnesota, Minneapolis, Minnesota 55455, USA}

\corres{Correspondence: kamenev@physics.umn.edu}

\abstract{We study dynamics and thermodynamics of ion channels, considered as effective 1D Coulomb systems.  The long range nature of the inter-ion interactions comes about due to the dielectric constants mismatch between the water and lipids, confining the electric filed to stay mostly within the water-filled channel.  Statistical mechanics of such Coulomb systems is dominated by entropic effects which may be accurately accounted for by mapping onto an effective quantum mechanics. In presence of multivalent ions the corresponding quantum mechanics appears to be non-Hermitian. In this review we discuss a framework for semiclassical calculations for corresponding non-Hermitian Hamiltonians. Non-Hermiticity elevates WKB action integrals from the real line to closed cycles on a complex  Riemann surfaces where direct calculations are not attainable. We circumvent this issue by applying tools from algebraic topology, such as the Picard-Fuchs equation. We discuss how its solutions relate to the thermodynamics and correlation functions of multivalent solutions within long water-filled channels.}


\begin{document}

\section{Introduction}\label{sec:Introduction}

Biological ion channels \cite{Stojilkovic2003} of cell membranes are vital parts of life. They allow for salt ions to pass through otherwise impermissible cell membranes. Their selectivity for specific ions regulates the concentrations inside the cell, which in turn keeps the cells functional \cite{MacKinnon}.  For  physicists they also provide a fascinating example of a quasi-1D statistical system formed by ions confined to move in a narrow water-filled tube inside a lipid membrane \cite{EdwardsLenard,PhysicaA,Demery,Kaufman2013,Kaufman2015,Kavokine,Experiment}. Another similar system is water-filled nanopores in silicon \cite{Li,Storm}. These are heavily used in genetic sequencing techniques where high-throughput of selective transport is the most important factor \cite{KTH}. Other examples include free-standing silicon nanowires \cite{Lieber1,Lieber} and water-filled nanotubes \cite{Takaiwa39,nanotubes1}. These systems play various different roles in biology and technology, however they follow the same underlying physics. What makes them special is the large ratio between the dielectric constants of water, $\kappa_1 \simeq 80$, and the surrounding media (e.g. for lipids  $\kappa_2\simeq 2$). Because of this, the electric field created by an ion within is confined to stay mostly inside the water-filled channel and does not leak into the lipid or other surrounding media. This simple observation has profound consequences.

First, as was noticed by Parsegian \cite{Parsegian}, it creates a potential barrier for an ion to enter the channel. This barrier is equal to the energy difference between an ion being inside and outside the channel.  For a channel of radius $a$ the electric field created by an ion of charge $e$ in the middle of the channel is $E_0=2e/(\kappa_1a^2)$. The corresponding field energy integrated over the channel volume is $U_0=\frac{\kappa_1}{8\pi}E_0^2 \pi a^2 L = k_BT (\lambda_B L)/(2a^2)$, where $L$ is the length of the channel and $\lambda_B=e^2/(\kappa_1 k_BT)\approx7$\AA~is the Bjerrum length at ambient temperature \cite{PhysicaA}. For a typical channel with $L\approx40$\AA~and $a\approx5$\AA~the corresponding (self-) energy barrier exceeds ambient temperature $k_BT$ by a factor of 5 or 6. This means that such a channel blocks the transport of ions. There are at least two mechanisms which nature employs to overcome this issue. One is placing charged radicals along the channel path. The other is entropic screening of the barrier by a collective effect of multiple positive and negative ions inside the channel. In this review we focus on this latter phenomena, while the former is addressed in Refs.~\cite{ZhangPRL,ZhangPRE,Kaufman2013,Kaufman2015}.

\begin{figure}
	\includegraphics[width=\textwidth]{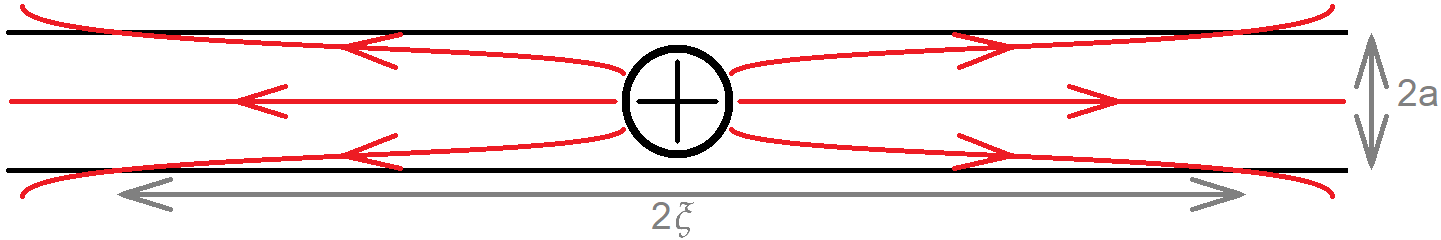}
	\caption{This is an illustration of the electric field lines emanating from an ion inside a water-filled channel of radius $a$ which is surrounded by a medium with lower dielectric constant. Due to the mismatch in dielectric constants the field lines run mostly along the channel which means that another charge would feel an effective 1D Coulomb potential. The ratio is finite however, i.e. a distance $\xi$ away from the ion the field lines start permeating the outside medium. If the channel is shorter than this critical length scale, $L<\xi$, or the typical spacing between charges is smaller than $\xi$, then all interactions are well-described by the 1D Coulomb potential.}
	\label{fig:Channel}
\end{figure}

The second consequence of the mismatch of dielectric constants is that the mutual interactions between the ions within the channel acquire the form of the 1D Coulomb potential
\begin{equation}\label{eq:Intro:potential}
	\Phi(x_i-x_j)=eE_0|x_i - x_j|, 
\end{equation}
where $x_i$ are 1D coordinates of the ions along the channel axis. As illustrated in Fig. \ref{fig:Channel}, the electric field lines emanating from a charge are bent to run along the channel. Only after a characteristic length $\xi$ given by the implicit relation $\xi^2=a^2\kappa_1/(2\kappa_2)\ln(2\xi/a)$ the field lines start penetrating the lipid membrane and escaping the channel \cite{PhysicaA}. For a water-filled channel in a lipid membrane this gives $\xi\approx7a$. Hence for a sufficiently short channel with $L<\xi$ or (as considered in Sec. \ref{sec:Large}) a large concentration of salt ions where the characteristic distance between two ions is smaller than $\xi$, the interactions follow an effective 1D Coulomb potential. The linear nature of the potential \eqref{eq:Intro:potential} leads to the curious observation that the energy barrier of transporting a charge through the channel can't be less than $U_0$, irrespective of how many other ions are present in the channel \cite{Parsegian}. Indeed, for the most favorable arrangement of alternating positive and negative ions, the electric field along the channel alternates between $\pm E_0$. This leads back to the value of $U_0$ for the electrostatic energy of adding a single ion to the channel in the presence of the other ions. This may seem as a predicament that collective screening can't lower the transport barrier. Such conclusion is premature, however. The resolution of this apparent paradox is that in a system of multiple particles at a finite temperature it is the {\em free energy} (rather than the {\em energy}) which determines the transport barrier. The difference between the two is given by the entropy, i.e. it is the entropy of the ion gas within the channel, which provides the screening mechanism. The nature of the entropic suppression of the transport barrier can be traced to the aforementioned independence of the energy $U_0$ of the positions of individual ions. This observation implies that there is a large number of microscopic configurations which are close in energy. This is the hallmark of a state with large entropy and thus lower free energy.

Formalizing  these observations is not entirely straightforward. As was first realized by Edwards and Lenard (EL) in 1962 \cite{EdwardsLenard} it requires mapping of the 1D statistical system onto an effective quantum mechanics with cosine potential.  In fact, this is a particular case of a generic correspondence between D-dimensional statistical mechanics of the Coulomb gas and (D-1)-dimensional sine-Gordon field theory \cite{AltlandSimons}. The D=2 version of this mapping is well-known in the physics of the Berezinskii-Kosterlitz-Thouless  transition. The less appreciated fact is that the Hermitian potential of the form $2\cos\theta=e^{i\theta} +e^{-i\theta}$  is a consequence of having a neutral plasma of ions with charges $\pm e$. Indeed, in  the EL mapping the $e^{\pm i\theta}$ operators shift the value of the electric field in the channel (the variable canonically conjugated to $\theta$)  by a quanta $\pm 2E_0$, which corresponds to the electric field generated by a unit charge $\pm e$.

What happens in the presence of a multivalent dissociated salt, such as e.g. $Ca Cl_2$ which produces a plasma with positive charges $+2e$ and twice as many negative charges $-e$?  It is not difficult to see that the EL mapping leads to an effective quantum mechanics with the potential 
$\frac{1}{2}e^{2i\theta}+e^{-i\theta}$. Such quantum mechanics in non-Hermitian and thus admits a complex-valued spectrum. This may present a problem for the interpretation of the original statistical mechanics of the Coulomb plasma. For example, the free energy density (a manifestly real quantity) is given by the logarithm of the partition function. Therefore the partition function needs to be real and positive. Fortunately the effective non-Hermitian quantum theories admit the so-called $\mathcal{PT}$-symmetry \cite{Bender2003}, which forces all eigenvalues to be real or appear as complex-conjugate pairs. When calculating the partition function, which includes summing over all eigenvalues, the imaginary parts cancel and we obtain a physical result \cite{JETP}. However in general there exist complex eigenvalues (spontaneously broken $\mathcal{PT}$-symmetry). This translates to an oscillatory character of certain correlation functions, reflecting short-range charge density wave correlations within the channel.

To model the transport of ions through the channel in this framework we use the concept of boundary charges which was developed in Ref. \cite{PhysicaA}. From now on we assume that the channel is sufficiently short so that all field lines stay inside the channel. If there are no ions inside the channel (or the sum of all charges is zero), then there is no electric field emanating from the channel. If a single ion is added in the center of the channel, then half of its electric field lines are exiting the channel on the left and the other half on the right, cf. figure \ref{fig:Channel}. This is akin to having two image boundary charges $q,q'=\frac{1}{2}$ at the two ends of the channel (charges are measures in units of $e$). These charges are provided by polarization effects in the well-conducting reservoirs. There are only integer charges inside the channel. Hence, if the boundary charge at one end is $q$ (the ion emits a fraction $q$ of its field lines at one end), then the other boundary charge is $q'=1-q$. Ref. \cite{PhysicaA} shows that moving a unit probe charge through the channel (while allowing the other ions to equilibrate) creates boundary charges which change from zero to one. Once the boundary charges reach an integer value they may either be released from the end points and join the bulk, or enter into the channel. This makes thermodynamic properties periodic functions of $q$ with unit period. In section \ref{sec:Model} we show that the boundary charge $q$ takes the role of the quasi-momentum in the effective quantum mechanics. Hence the bandwidth of the lowest quantum-mechanical band translates directly to the transport barrier.

This review is devoted to the mathematical apparatus needed to treat the non-Hermitian quantum mechanics appearing in physics of multivalent 1D plasmas. However, these methods can be applied more broadly to a wide range of non-Hermitian systems. In particular we focus on semiclassical methods applicable for relatively large concentrations of the dissociated salts. Our central observation is that the corresponding (complex) semiclassical trajectories may be viewed as closed cycles on Riemann surfaces of non-zero genus. The action integrals along such cycles are given by solutions of the Picard-Fuchs differential equation, allowing for their analytic evaluation. As a result, one obtains asymptotically exact thermodynamic and correlation functions of the 1D multivalent Coulomb plasmas. Of particular interest is the transport barrier, given by the width of the lowest Bloch band (i.e. energy difference between anti-periodic and periodic ground-states of the Schr\"odinger equation). We obtain analytic results for the transport barriers for various combinations of ion valencies as functions of salt concentration and temperature.

The structure of this paper is as follows: in Section \ref{sec:Model} we discuss the EL mapping to cosine quantum mechanics and its generalizations in the non-Hermitian cases. Section \ref{sec:Large} is devoted to the semiclassical treatment of the corresponding non-Hermitian theories using the Picard-Fuchs equation. At the end of that section we go beyond the usual semiclassical formulas and describe how to obtain second- and higher-order corrections with little computational effort. We provide a brief summary and discussions in Section \ref{sec:Summary}.

\section{Thermodynamic description and equivalent quantum mechanics}
\label{sec:Model}

In this section we discuss the relationship between statistical mechanics of the ion channel and (non-Hermitian) quantum mechanics. We start with a thermodynamic description of the ion channel in terms of the grand-canonical partition function. Then we review how to map the partition function onto a Feynman propagator, and derive a Hamilton operator from there. This mapping was pioneered by Edwards and Lenard \cite{EdwardsLenard}, and subsequently used by several authors as starting point \cite{PhysicaA,JETP,JoPA,CUMS}. If the system consists of anions and cations with the same valency and concentration, then the resulting Hamilton operator is Hermitian. However, if the positive and negative charges have different valency, for example solutions of the divalent salts $MgCl_2$ or $CaCl_2$, non-Hermitian terms appear. Hence the spectrum of the resulting operator also contains complex eigenvalues. We discuss how reality and positivity of the partition function is ensured. In the end we comment on the case if charge neutrality is violated.

\subsection{Derivation of the Hamilton operator}
As discussed in section \ref{sec:Introduction} charged ions inside the channel interact with the effective 1-dimensional Coulomb potential $\Phi(x) = -eE_0|x|$, where $E_0=2e/\kappa_1a^2$ is the electric field strength from a single ion with charge $e$ inside a channel with radius $a$ and dielectric constant $\kappa_1$ \cite{PhysicaA}. The total interaction energy of all ions in the channel is given by
\begin{equation}\label{eq:Model:Interaction}
	U = \frac{1}{2} \iint_0^L dxdx' \rho(x)\Phi(x-x')\rho(x').
\end{equation}
Here we write the charge density for point charges in terms of $\delta$-functions:
\begin{equation}\label{eq:Model:ChargeDensity}
	\rho(x) = \sum_{j=1}^{N_1+N_2} \sigma_j\delta(x-x_j) + q\left(\delta(x) - \delta(x-L)\right),
\end{equation}
where $\sigma_j=n_1$ for $1\leq j \leq N_1$ and  $\sigma_j=-n_2$ for $N_1+1\leq j \leq N_1+N_2$. This charge density represents $N_1$ cations with valency $n_1$ and $N_2$ anions with valency $-n_2$, and the two fractional boundary charges $\pm q$ at $x=0,L$. The channel is open and can exchange particles with two 3D bulk reservoirs at the ends. Therefore the thermodynamic properties are given by the grandcanonical partition function,
\begin{equation}\label{eq:Model:PartitionAnsatz}
	\mathcal{Z} = \sum_{N_1,N_2=0}^\infty \frac{f_1^{N_1}f_2^{N_2}}{N_1!N_2!} \prod_{j=1}^{N_1+N_2}
	\int_0^L dx_j e^{-U/k_BT},
\end{equation}
where $f_{1,2}$ are the fugacities of the two charge species. As shown in Refs.~\cite{EdwardsLenard,PhysicaA}, the partition function can be converted into a functional integral by introducing an auxiliary field $\theta(x)$ as conjugate to the charge density $\rho(x)$. Through this process all $x_j$ integrals decouple, bringing them to the form $\sum_N [f\int\! dx\, e^{i\sigma \theta(x)}]^N/N!=\exp\{f\int\! dx\, e^{i\sigma \theta(x)}\}$. The interaction potential \eqref{eq:Model:Interaction}, being inverse of the 1D Laplace operator, leads to an additional term $\exp\{(k_BT/eE_0)\int\! dx\,  \theta \partial_x^2 \theta\}$. As a result the partition function \eqref{eq:Model:PartitionAnsatz} is identically written in terms of the Feynman path integral with an "imaginary time" $x$  for quantum mechanics with the Hamiltonian
\begin{equation}\label{eq:Model:Hamiltonian}
	\hat{H}=(i\partial_\theta-q)^2-\left(\alpha_1 e^{in_1\theta}+\alpha_2 e^{-in_2\theta} \right)\, ,
\end{equation}
where $\alpha_{1,2}=f_{1,2}k_BT/eE_0$ are dimensionless ion concentrations. The Feynman integral is the expectation value of the evolution operator during imaginary ``time'' $L$,
\begin{equation}\label{eq:Model:Partition}
	\mathcal{Z}_L=\left\langle q\Big|{\cal X} e^{-\frac{eE_0}{k_BT}\int_0^L\! dx\, \hat{H} } \Big|q\right\rangle
	=\sum_m |\langle q|m\rangle|^2  e^{-\frac{eE_0L}{k_BT} \varepsilon_m(q)},
\end{equation}
where ${\cal X}$ stands for $x$-ordered exponent. Here $\{\varepsilon_m(q)\}_m$ is the spectrum of the effective Hamiltonian $\hat{H}$, and $|m\rangle =\psi_m(\theta)$ are its eigenvectors in the Hilbert space of periodic functions $\psi_m(\theta)=\psi_m(\theta+2\pi)$. The matrix elements are $ \langle q|m\rangle= \int_0^{2\pi}\!\! d\theta e^{-iq\theta} \psi_m(\theta)$. The boundary charge $q$ plays the role of the Bloch quasi-momentum and the spectrum is periodic in $q$ with unit period.

Note that for $\alpha_1=\alpha_2$ and $n_1=n_2$ the potential in equation \eqref{eq:Model:Hamiltonian} reduces to a cosine function and the Hamiltonian becomes the well-known Mathieu Hamiltonian \cite{EdwardsLenard}. However, if these conditions are violated the potential is non-Hermitian \cite{JETP}. We discuss implications of this in the next section.

\subsection{Physical observables}
The partition function in equation \eqref{eq:Model:Partition} gives the thermodynamic properties of the ion gas. However, to be physically meaningful the partition function needs to be real and positive, while the spectrum of the non-Hermitian Hamiltonian \eqref{eq:Model:Hamiltonian} may contain non-real eigenvalues. This issue is resolved because the Hamiltonian obeys a symmetry akin to $\mathcal{PT}$-symmetry. The combined action of the "parity operator" $\mathcal{P}: \theta\to-\theta$ and "time reversal" $\mathcal{T}: i\to-i$ leaves the Hamiltonian in equation \eqref{eq:Model:Hamiltonian} unchanged. Bender et al \cite{Bender2003} proved that all eigenvalues of a $\mathcal{PT}$-symmetric Hamiltonian are either real or appear in complex conjugated pairs. Hence, summing over all eigenvalues in equation \eqref{eq:Model:Partition} gives a real result. In \cite{JETP} is was shown that for positive values of concentrations $\alpha_{1,2}>0$ the lowest energy band $\varepsilon_0(q)$ is entirely real, ensuring positivity of the partition function. The higher bands $\varepsilon_m(q)$ are in general complex-valued.

Hence we obtain a physically meaningful partition function, and can connect it to thermodynamic observables. The pressure of the Coulomb gas is its free energy per unit length
\begin{equation}\label{eq:Model:Pressure}
	P= k_B T \frac{\partial \ln \mathcal{Z}_L}{\partial L} \,  \stackrel{L\to \infty}{\longrightarrow}\, -eE_0 \varepsilon_0(q)\,,
\end{equation}
which for a long channel is determined by the eigenvalue with the smallest real part, $\varepsilon_0(q)$. In equilibrium the system minimizes its free energy by choosing an appropriate boundary charge $q$. In \cite{JETP,JoPA} this minimum was found to generally be the non-polarized state of the channel, i.e. $q=0$. Adiabatic charge transfer through the channel is associated with the boundary charge $q$ sweeping through its full period. As a result, the (free) energy barrier for ion transport is
\begin{equation}\label{eq:Model:Barrier}
	U_0=eE_0 L \Delta_0\,,
\end{equation}
where $\Delta_0$ is the width of the lowest Bloch band. Therefore the ground state energy and the width of the lowest Bloch band of the Hamiltonian \eqref{eq:Model:Hamiltonian} give the leading thermodynamic and transport properties of the $(n_1,n_2)$ Coulomb gas. In section \ref{sec:Large} we discuss analytic results for the eigenvalues and the bandwidth.

\subsection{Charge non-neutrality}
In \cite{EdwardsLenard} it was shown that for arbitrary values of $\alpha_{1,2}$ the Hamiltonian \eqref{eq:Model:Hamiltonian} is always isospectral to a similar charge-neutral Hamiltonian. This can be seen by shifting the coordinate as $\theta\to\theta+\theta_0$. Upon such transformation the dimensionless concentrations $\alpha_{1,2}$ renormalize as $\alpha_1\to \alpha_1 e^{in_1\theta_0}$ and $\alpha_2\to \alpha_2 e^{-in_2\theta_0}$. Notice that the combination $\alpha_1^{n_2}\alpha_2^{n_1}$ remains invariant. Hence the family of Hamiltonians \eqref{eq:Model:Hamiltonian} with
\begin{equation}\label{eq:Model:Isospectral}
	\alpha_1^{n_2}\alpha_2^{n_1} = {\rm const}
\end{equation}
is isospectral \cite{EdwardsLenard,JETP}. Therefore one may choose one representative from each isospectral family. A convenient choice is taking the representative with charge neutrality in the bulk reservoirs, i.e. $n_1\alpha_1=n_2\alpha_2\equiv\alpha$. The physical reason for this symmetry is that the interior region of the channel always preserves charge neutrality due to the large self-energy of charges. The edge regions screen charge imbalances of the reservoirs. Therefore, irrespective of the relative fugacities of cations and anions in the reservoirs, the thermodynamics of the long channel are equivalent to the one in contact with neutral reservoirs with an appropriate salt concentration $\alpha$. This brings the Hamiltonian \eqref{eq:Model:Hamiltonian} to the form 
\begin{equation}\label{eq:Model:HamiltonianIso}
	\hat{H} =\alpha\left[ \hat p^2-  \left(\frac{1}{n_1}\,e^{in_1\theta}+\frac{1}{n_2}\, e^{-in_2\theta}\right)\right]\,.
\end{equation}
where  we define the momentum operator as
\begin{equation}\label{eq:Model:Momentum}
	\hat p = \alpha^{-1/2}(-i\partial_\theta +q)\,; \quad\quad [\theta, \hat p]=i\alpha^{-1/2}\,.
\end{equation}
The commutation relation shows that $\alpha^{-1/2}$ plays the role of the effective Planck constant. Hence a large concentration of charges corresponds to the semiclassical limit of the Hamiltonian \eqref{eq:Model:HamiltonianIso}. We further rescale the eigenvalues $\varepsilon$ as
\begin{equation}\label{eq:Model:RescaledEnergy}
	u = \frac{n_1n_2}{n_1+n_2}\frac{\varepsilon}{\alpha}.
\end{equation}
This keeps the classical minimum of the potential at $u=-1$, irrespective of the concentration $\alpha$ and the valencies $n_1,n_2$. In section \ref{sec:Large} we discuss the spectral properties of the Hamiltonian \eqref{eq:Model:HamiltonianIso} in the semiclassical limit.

\section{Large charge concentration}
\label{sec:Large}

In section \ref{sec:Model} we mapped the grand-canonical partition function of the Coulomb gas onto an equivalent quantum system. The resulting Hamiltonian, equation \eqref{eq:Model:HamiltonianIso}, contains one free parameter $\alpha$ which is proportional to the concentration of charged ions. In this section we analyze the spectral problem of this Hamiltonian in the limit of large charge concentration $\alpha$. As argued after equation \eqref{eq:Model:Momentum}, this is the semiclassical limit of the equivalent quantum problem. We use the main semiclassical results, Bohr-Sommerfeld quantization and Gamow's formula, to calculate the eigenvalues and bandwidths of the Hamiltonian for several different cases of the valencies $(n_1,n_2)$. In the case of equal valencies, $n_1=n_2$, the Hamiltonian \eqref{eq:Model:HamiltonianIso} is the well-known Mathieu Hamiltonian which we discuss in section \ref{sec:Large:Equal}. It's spectral properties were calculated using several different approaches \cite{EdwardsLenard,CUMS,PhysicaA,JETP,JoPA,thesisPhD}. In this review we focus on an approach based on integration on a complex Riemann surface \cite{JETP,JoPA,thesisPhD}. We choose this method because it can also be applied to the cases with different valencies, $n_1 \neq n_2$, see section \ref{sec:Large:Different}. In that situation the Hamiltonian is non-Hermitian, and the required action integrals are not attainable by straightforward integration. Instead we show how to relate them to integrals along closed cycles on a Riemann surface. Then we use powerful tools from algebraic topology to derive a differential equation for the action integrals. This is known as the Picard-Fuchs equation. The required actions are a combination of the solutions of this differential equation. Through this procedure we bypass the use of  direct integration methods. From the actions we obtain the eigenvalues and the bandwidths, which are directly related to the pressure and the transport barrier of the ion channel. In section \ref{sec:Large:Secondorder} we go one step further. We use the same concepts to calculate the second-order corrections in the WKB series. Most importantly we show that these can be expressed in terms of the already-calculated action and its derivatives, and therefore can be obtained with minimal computational effort. This gives an improved semiclassical approximation of the eigenvalues. Relating this to the pressure in the ion channel we find that beyond the ideal-gas pressure and the Debye-Hueckel correction there is another correction which only depends on the geometry of the channel but not on the concentration of ions. We compare these results to numerical calculations.

\subsection{Equal valency}\label{sec:Large:Equal}
As mentioned in section \ref{sec:Model} the Hamiltonian in equation \eqref{eq:Model:Hamiltonian} is Hermitian if the valencies of the two charges are equal, $n_1=n_2$. Indeed, in this case it reduces to the well-known Mathieu Hamiltonian,
\begin{equation}\label{eq:Large:HamEqual}
	\mathcal{H} = \alpha\left[ \hat{p}^2 - 2\cos\theta \right].
\end{equation}
In literature there exist several studies of the Coulomb gas with charges of equal valency. In \cite{EdwardsLenard} it was first noted that the Coulomb gas is mapped onto the Mathieu equation. \cite{CUMS} performs a semiclassical calculation on this equation via direct integration. From this they obtain the required actions and analytic approximations of the eigenvalues and bandwidths. \cite{PhysicaA} provides additional qualitative arguments which lead to the same results. However, as mentioned above, in this section we will follow the Riemann surface methods developed in \cite{JETP} because in that framework one can also study the case of unequal valencies $n_1\neq n_2$ in section \ref{sec:Large:Different}, and these concepts form the basis of our considerations for higher-order corrections in section \ref{sec:Large:Secondorder}.

\subsubsection{Construction of the Riemann surface}

\begin{figure}[h!]
	\fbox{\includegraphics[width=\textwidth]{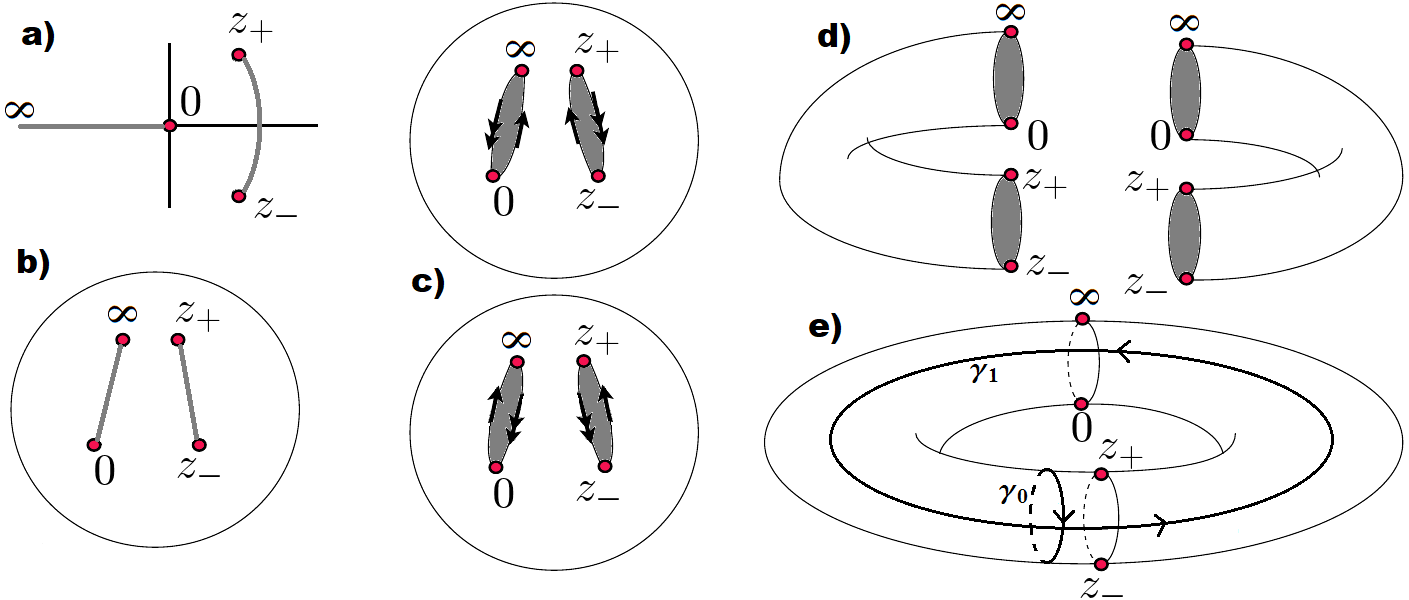}}
	\caption{Construction of the Riemann surface of genus 1, as defined by equation \eqref{eq:Large:Riemann11}. (a) In the $z$-plane there are four branch points at $0,z_\pm,\infty$ which are pairwise connected by two branch cuts (gray). (b) Considering $z=\infty$ as a regular point the complex plane compactifies to a Riemann sphere with two cuts on the sphere. (c) The double-valued nature of the function $p(z)$ is resolved by defining two copies of the Riemann sphere. The branch cuts are opened and the spheres are deformed into tubes (d) and glued together to form a torus (e). The arrows are used to signify the edges that are glued together. There are two fundamental cycles $\gamma_0,\gamma_1$ which are topologically different and non-trivial, i.e. they can not be smoothly transformed into each other or a point. Reproduced with permission from Ref. \cite{JoPA}.}
	\label{fig:TorusConstruction}
\end{figure}

In the semiclassical ansatz we look for wave functions of the form $\psi = e^{i\alpha^{1/2}S}$, where $S$ is an action for the classical problem with the normalized Hamiltonian \eqref{eq:Large:HamEqual}. The semiclassical trajectories satisfy classical Hamiltonian equations of motion and thus conserve the (complex)  energy $u$ in equation \eqref{eq:Model:RescaledEnergy},
\begin{equation}\label{eq:Large:u11}
	2u= p^2-2 \cos \theta \,.
\end{equation}
In this normalization $u=\mp 1$ corresponds to the bottom (top) of the cosine potential. Our approach to calculate the action integrals $S = \oint_\gamma p(\theta,u) d\theta$ is based on complex algebraic topology. First we set $z=e^{i\theta}$ and consider $(z,p)$ as complex variables. The energy conservation (\ref{eq:Large:u11}) defines thus a family of complex algebraic curves, parametrized by $u$ and satisfying 
\begin{equation}\label{eq:Large:Riemann11}
	\mathcal{E}_u:\quad\quad {\cal F}(p,z)=p^2 z - (z^2+2uz+1) = 0.
\end{equation}
For $u\neq\pm 1$ it can be checked that $(\partial {\cal F}/\partial z,\partial {\cal F}/\partial p)$ does not vanish on ${\cal E}_u$, so each $\mathcal{E}_u$ is nonsingular. Then ${\cal F}(p,z)$ implicitly defines a locally holomorphic map $p=p(z)$. The exceptions to this occur at $z=0,\infty, z_\pm$, where $z_\pm = -u \pm i\sqrt{1 - u^2}$ are the roots of $p^2=0$ (i.e. classical turning points). In a vicinity of these four branch points  $p(z)$ behaves as
\begin{align}\label{eq:Lage:BranchPoints11}
	&p\sim z^{-1/2},& (z\sim 0)\\\nonumber
	&p\sim z^{1/2},& (z\sim \infty)\\\nonumber
	&p\sim (z-z_\pm)^{1/2},& (z\sim z_\pm)
\end{align}
respectively, i.e. $p(z)$ is locally double-valued. Note that we added the point at infinity to have an even number of branch points, which compactifies the complex plane and makes it topologically equivalent to a Riemann sphere, cf. figure \ref{fig:TorusConstruction}. 

To avoid dealing with the double-valued function, $p(z)$, we introduce a second copy of the complex $z$-plane and the corresponding Riemann sphere. On both sheets we define two branch cuts connecting the four branch points, between $0,\infty$ and the turning points $z_\pm$ respectively. $p(z)$ is analytically continued across the branch cut, i.e. when crossing a branch cut we jump from the first sheet to the second and vice versa. Identifying the branch cuts as edges we can deform the two Riemann spheres into tubes and glue them together to form a torus. This construction is visualized in figure \ref{fig:TorusConstruction}. Thus the complex algebraic curve $\mathcal{E}_u$ in equation \eqref{eq:Large:Riemann11} defines a torus which is a compact Riemann surface of genus $g=1$. (Generically, every compact Riemann surface is topologically equivalent to a sphere with some number of handles $g$, or a (multi-)torus with $g$ holes, called the genus of the surface).

\subsubsection{Integrals on the Riemann surface and the Picard-Fuchs equation}
The action integrals can be understood  as $S(u)=\oint_\gamma \lambda(u)$ over closed cycles $\gamma$, where
\begin{equation}\label{eq:Large:1form11}
	\lambda(u)=p(\theta)\,d\theta=p(z)\frac{dz}{iz}=\frac{(z^2+2uz+1)^{1/2}}{iz^{3/2}}\, dz
\end{equation}
is the action 1-form which, by construction, is holomorphic on the Riemann surface. 

\begin{figure}[b!]
	\includegraphics[width=0.5\textwidth]{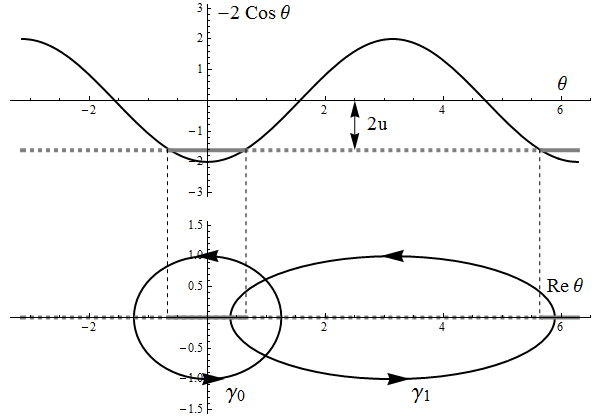}\hspace{0.05\textwidth}
	\includegraphics[width=0.35\textwidth]{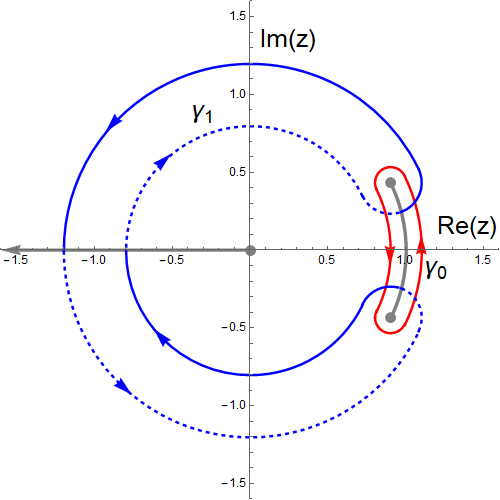}
	\caption{Left: The classically allowed (forbidden) regions along the $\theta$-axis at energy $u$ are shown by the solid (dashed) gray line. Deforming the classical (instanton) orbits into the complex plane leads to the cycles $\gamma_0 (\gamma_1)$. \\
	Right: Cycles $\gamma_0$ (red) and $\gamma_1$ (blue) on the complex $z$-plane for $u=-0.9$. Notice that the cycle $\gamma_1$ crosses the two cuts from the first sheet (solid line) to the second sheet (dashed  line) and back. Reproduced with permission from Ref. \cite{JETP}.}
	\label{fig:Cycles11}
\end{figure}

To visualize the relevant trajectories we momentarily return to $\theta$ and consider it as complex. In this representation one has square-root branch cuts along the real axis, connecting the classical turning points. The action integrals run just above or below the real axis in between the turning points. Combining them into  closed cycles, one can push these cycles off the real axis and away from the turning points without altering the integrals (by Cauchy's theorem). We call these the classical cycle $\gamma_0$ and the instanton cycle $\gamma_1$ , shown in figure \ref{fig:Cycles11}. Translating these two cycles to the complex $z$-plane yields the contours in the right panel of figure \ref{fig:Cycles11}.

Cauchy's theorem is also valid on the Riemann surface since the action form (\ref{eq:Large:1form11}) is, by construction, holomorphic on the torus. Therefore all closed cycles can be deformed without changing the integrals, and can be expressed as a combination of an integer number of these two basis cycles. This leads to our key idea how to calculate the action integrals: for this we employ a central theorem of algebraic topology, de Rham's theorem. It states that on a Riemann surface there are exactly as many linearly independent holomorphic 1-forms to integrate upon as there are independent closed cycles to integrate along. This is valid up to exact forms, i.e. 1-forms which integrate to 0 along any closed cycle, and boundaries, i.e. closed curves which can be continuously deformed to a point. Hence there are exactly two independent holomorphic 1-forms on the Riemann surface. Any set of three 1-forms is linearly dependent modulo an exact form which integrates to 0 upon integration along any closed cycle.\footnote{A full explanation of the mathematical concepts is beyond the scope of this review. A detailed discussion of relevant and related concepts is in \cite{Miranda}, basic definitions and additional background are in \cite{doCarmo,Ahlfors}. All concepts can also be found online at \cite{EncyclopediaMath}. A simplified derivation specifically for complex-valued Riemann surfaces is in chapter 2 of \cite{thesisPhD}.}

Equipped with this we look at a set which contains the action 1-form \eqref{eq:Large:1form11} and its first two derivatives with respect to energy $u$, $\{\lambda(u), \lambda'(u), \lambda''(u)\}$. Taking derivatives does not change the structure of branch points, therefore these are three 1-forms which are all defined on the same Riemann surface. Hence we know that there must exist a linear combination of these which is an exact form. Ref. \cite{JETP} explains in detail how to find the linear combination and the exact form as
\begin{equation}\label{eq:Large:ExactForm11}
	\left((u^2-1)\partial_u^2+\frac{1}{4}\right)\lambda(u) = \frac{d}{dz}\left[\frac{i}{2}\, \frac{1-z^2}{z^{1/2}(z^2+2uz+1)^{1/2}}\right]dz\,.
\end{equation} 
It is evident from Stokes' theorem that the right-hand-side integrates to 0 along any closed cycle on the Riemann surface. Hence we obtain
\begin{equation}\label{eq:Large:PF11}
	\oint_{\gamma} \left((u^2-1)\partial_u^2+\frac{1}{4}\right)\lambda(u) = (u^2-1)S^{\prime\prime}(u) + \frac{1}{4}\,S(u) = 0\,.
\end{equation}
This differential equation for the action $S(u)$ is called the Picard-Fuchs equation \cite{Miranda}. Integration is performed along a closed cycle $\gamma$, which can be the classical or the instanton cycle, $\gamma_{0,1}$ in figure \ref{fig:Cycles11}. Therefore both the classical and instanton actions $S_{0,1}(u)$ are solutions of the Picard-Fuchs equation \eqref{eq:Large:PF11}. This equation is a second-order ordinary differential equation, therefore it admits two independent solutions. These can be found in the form $F_0(u^2)$ and $u F_1(u^2)$, where
\begin{align}\label{eq:Large:Solutions11}
	F_0(u^2)&= \,_2F_1\left(-\frac{1}{4},-\frac{1}{4};\frac{1}{2};\,u^2\right),
	\\\nonumber
	F_1(u^2)&= \,_2F_1\left(+\frac{1}{4},+\frac{1}{4};\frac{3}{2};\, u^2\right),
\end{align}
are hypergeometric functions \cite{Heckman,Mathworld}. These solutions form a basis out of which $S_{0,1}(u)$ must be composed, so we write
\begin{align}\label{eq:Large:Combination11}
	S_0(u)&=C_{00} F_0(u^2)+C_{01} u F_1(u^2),
	\\\nonumber
	S_1(u)&=C_{10} F_0(u^2)+C_{11} u F_1(u^2).
\end{align}
To find the correct coefficients $C_{jk}$, $j,k=0,1$ it is sufficient to evaluate the periods at one specific value of $u$. Employing the fact that the hypergeometric functions \eqref{eq:Large:Solutions11} are normalized and analytic at $u=0$, i.e. $F_k(u^2)=1+{\cal O}(u^2)$, one notices that $S_j(u)=C_{j0}+u C_{j1}+{\cal O}(u^2)$. Thus to identify $C_{jk}$ we expand the integrand $\lambda(u)$ to first order in $u$ and evaluate the integrals $S_j(u)$ at $u=0$. Straightforward calculation yields
\begin{align}\label{eq:Large:Coefficients11}
	&C_{00}=e^{-i\pi/2}C_{10}=8\pi^{-1/2}\Gamma(3/4)^2,
	\\\nonumber
	&C_{01}=e^{+i\pi/2}C_{11}=\pi^{-1/2}\Gamma(1/4)^2.
\end{align}
The relations between $C_{0k}$ and $C_{1k}$ are not accidental. They originate from the fact that the cycle $\gamma_1$ transforms into $\gamma_0$ by substitution $z'=e^{-i\pi}z$ and $u'=e^{i\pi}u$, and vice versa. This gives a global symmetry between the two periods,
\begin{eqnarray}\label{eq:Large:Symmetry11}
	S_0(u)=e^{-i\pi/2}S_1(e^{i\pi}u)\,.
\end{eqnarray}
Equations \eqref{eq:Large:Solutions11}--\eqref{eq:Large:Symmetry11} fully determine the classical and instanton actions $S_{0,1}(u)$. We now proceed to relate them to physical observables.

\subsubsection{Semiclassical results}
We seek semiclassical results for the sequence of low-energy bands terminated at $u=-1$. Therefore we quantize the classical action $S_0(u)$ according to the Bohr-Sommerfeld rule to determine the normalized energies $u_m$ as solutions of the equation
\begin{equation}\label{eq:Large:BS11}
	S_0(u_m)=2\pi \alpha^{-1/2}(m+1/2)\,,\quad\quad  m=0,1,\ldots
\end{equation}
We see that the cycle $\gamma_0$ contracts to a point when the energy goes to the bottom of the potential, $u\to-1$. This corresponds to vanishing of the classical action, $S_0(u=-1)=0$. To obtain an approximate analytic expression for the lowest energy levels $\varepsilon_m=2\alpha u_m$ we expand the classical action to first order near the bottom of the potential, 
\begin{equation}\label{eq:Large:Expand11}
	S_0(u) = 2\pi(u+1).
\end{equation}
Equations \eqref{eq:Large:BS11} and \eqref{eq:Large:Expand11} combined imply $\varepsilon_m = -2\alpha+2\alpha^{1/2}\left(m+1/2\right)$. As a result the pressure \eqref{eq:Model:Pressure} of a monovalent gas is
\begin{equation}
	P = -eE_0\varepsilon_0 = 2k_BT f - \sqrt{k_BTeE_0 f}.
\end{equation}
The two terms here are the pressure of the ideal gas with fugacity $f$ and the mean-field Debye-Hueckel interaction correction \cite{ZhangPRE}.

The instanton action $S_1(u)$ determines the bandwidth $(\Delta u)_m$ according to Gamow's formula,
\begin{equation}\label{eq:Large:Gamow11}
	(\Delta u)_m=\frac{\omega}{\pi\sqrt{\alpha}}\, e^{i\alpha^{1/2} S_1(u_m)/2}\,.
\end{equation}
Here $\omega=2$ is the frequency of the harmonic-oscillator approximation of the potential near the classical minimum. We expand the instanton action near the classical minimum and at the quantized energies $u_m=-1+\alpha^{-1/2}(m+1/2)$ we obtain
\begin{equation}\label{eq:Large:Instanton11}
	S_1(u_m) = 16i+2i\left(m+\frac{1}{2}\right)\ln\left(\frac{m+1/2}{32e\alpha^{1/2}}\right)\,.
\end{equation}
Applying this to the Gamow formula \eqref{eq:Large:Gamow11} leads to
\begin{equation}\label{eq:Large:Bandwidth11}
	(\Delta\varepsilon)_m = 2\alpha(\Delta u)_m =\frac{4}{\pi}\left(\frac{32e}{m+1/2}\right)^{m+1/2}\, e^{-8\alpha^{1/2}+(m/2+3/4)\ln\alpha},
\end{equation}
This coincides with the known asymptotic results for the Mathieu equation \cite{CUMS,MeixnerSchaefke,AbramowitzStegun}. As explained below equation (\ref{eq:Model:ChargeDensity}), adiabatic charge transport is associated with a change of the boundary charge $q$ (i.e. quasi-momentum) across the interval $0<q<1$ (i.e. the Brillouin zone). Therefore the {\em free energy} transport barrier is given by the width of the lowest Bloch band, $(\Delta\varepsilon)_0$. One notices that increasing the concentration of salt ions leads to an exponential entropic suppression of the transport barrier, $(\Delta\varepsilon)_0\propto \alpha^{3/4} e^{-8\sqrt{\alpha}}$.

\subsection{Multivalent ions}\label{sec:Large:Different}
So far we worked with the Hermitian example of the Mathieu Hamiltonian, i.e. when both ion species are monovalent, $n_1=n_2=1$. With that we could validate the Riemann surface method by comparing the results to literature. In this section we discuss cases of multivalent ions, we consider $n_1>n_2$ without loss of generality. In such a scenario the Hamiltonian \eqref{eq:Model:Hamiltonian} is non-Hermitian. This leads to complex values in the spectrum, which we present in section \ref{sec:Different:Spectrum}. Furthermore, in classical motion the coordinate and momentum acquire complex values. This results in a phase space $(\theta,p)$ with two complex dimensions (instead of two real dimensions). The classical (instanton) action is obtained by integrating the momentum $p(\theta)$ along the trajectory which connects two turning points and solves the classical equations of motion with real (imaginary) time. However, solving the equations of motion in complex phase space $(\theta,p)$ is non-trivial, if at all attainable. Therefore we go from an integral along the trajectory to an integral along a closed cycle in the plane of complex $z=e^{i\theta}$ which encloses the trajectory, similar to the mapping in figure \ref{fig:Cycles11}. With that we connect the non-Hermitian problem to the method that we validated in the previous section. We discuss this calculation for four different combinations of charge valencies in section \ref{sec:Different:PF}. In section \ref{sec:Different:Semiclassical} we connect the results to the classical and instanton actions.

\subsubsection{Spectrum of the non-Hermitian Hamiltonian}\label{sec:Different:Spectrum}

\begin{figure}[b!]
	\includegraphics[width=0.45\textwidth]{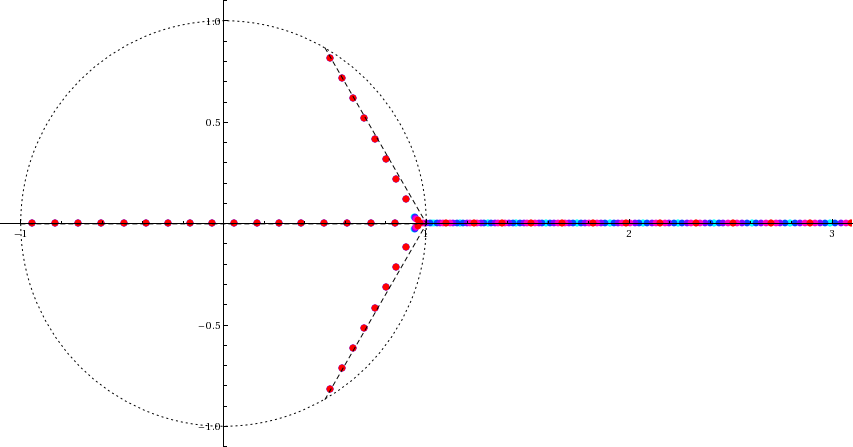}\vspace{0.05\textwidth}
	\includegraphics[width=0.45\textwidth]{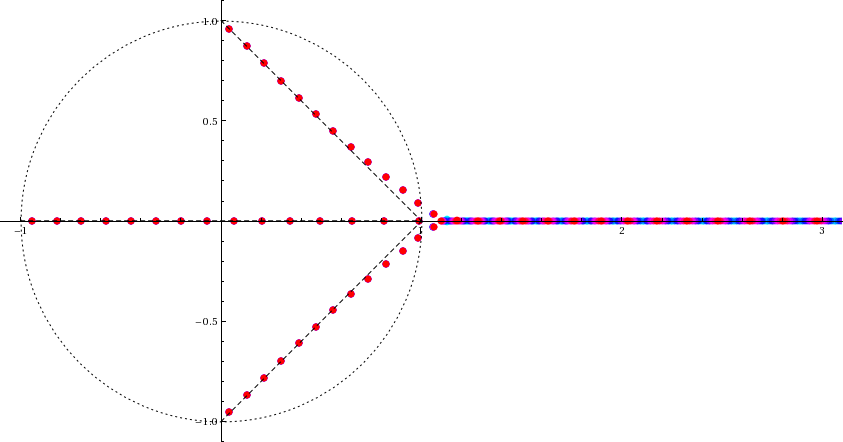}\\
	\includegraphics[width=0.45\textwidth]{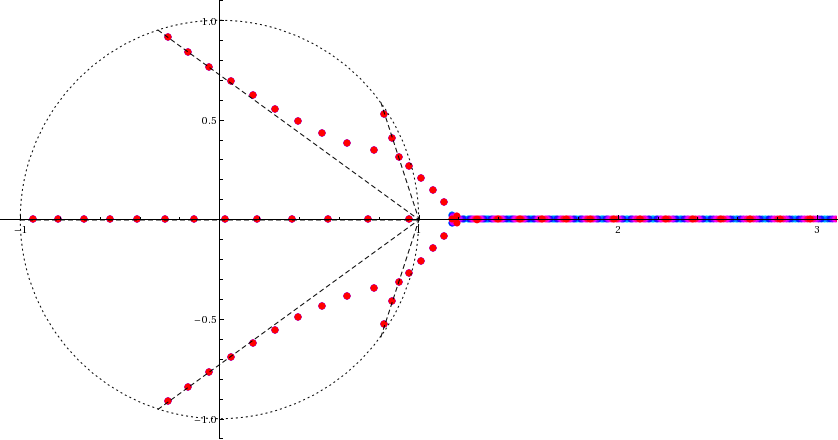}\vspace{0.05\textwidth}
	\includegraphics[width=0.45\textwidth]{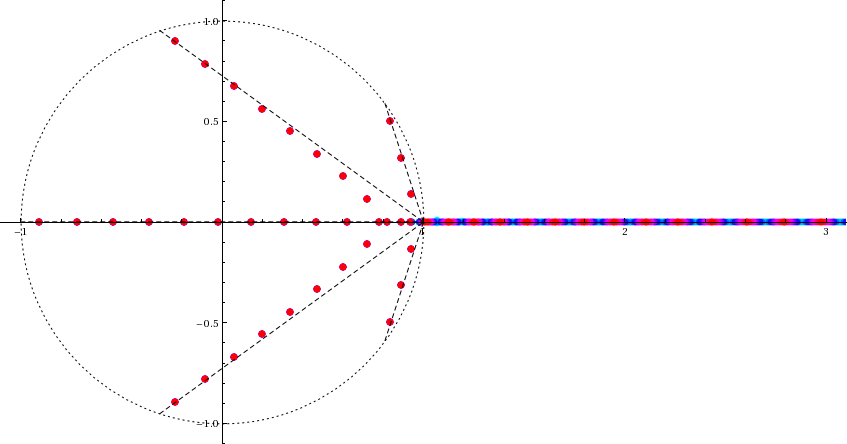}
	\caption{The bands of the non-Hermitian Hamiltonian in space of complex energy $u$. Blue stands for $q=0$, while red stands for $q=\frac{1}{2}$. The dotted circle marks $|u|=1$. In all cases we see multiple branches of narrow bands with complex values which terminate near the unit circle. The dashed line is a guide to the eye which connects the termination points of the branches, $u=-(1)^{1/(n_1+n_2)}$, to $u=1$. Top left: $(n_1,n_2)=(2,1)$, $\alpha=200$; top right: $(3,1)$, $\alpha=300$; bottom left: $(4,1)$, $\alpha=400$; bottom right: $(3,2)$, $\alpha=400$. Reproduced with permission from Refs. \cite{JETP,JoPA}.}
	\label{fig:Spectrum}
\end{figure}

Non-Hermiticity of the Hamiltonian \eqref{eq:Model:HamiltonianIso} has a significant effect on its spectrum. Namely, not all eigenvalues are real. In figure \ref{fig:Spectrum} we show numerical results for the eigenvalues at large concentration $\alpha$, for four different cases of the integers $(n_1,n_2)$. Most importantly all non-real eigenvalues appear as complex conjugate pairs. This is a consequence of the $\mathcal{PT}$-symmetry of the Hamiltonian and crucial to obtain a physically meaningful partition function, as discussed in section \ref{sec:Model}. Furthermore we see sequences of narrow bands which emerge from $u=-\nu$ with $\nu^{n_1+n_2}=1$. These sequences approximately follow the lines connecting $u=-\nu$ and $u=1$, but avoid the special point $u=1$. At some point all of these branches merge. Beyond this the nature of the spectrum changes drastically, instead of narrow bands and large gaps we see wide bands separated by small gaps. This feature is similar to the case of a periodic Hermitian potential: as long as the energy lies below the maximum of the potential there are narrow bands, while for energies exceeding the maximum there are wide bands. Hence we associate the point where the spectral branches meet with the top of the potential.\footnote{It is important to bear in mind that for a complex-valued potential there is no proper definition of a "maximum".} The energy variable $u$ is normalized so that in the Hermitian $(1,1)$ case this point lies at $u=1$. In the non-Hermitian cases we observe $u\approx0.96$ for $(2,1)$, $u\approx1.09$ for $(3,1)$, $u\approx1.20$ for $(4,1)$, and $u\approx0.84$ for $(3,2)$. These values are independent of $\alpha$, so this is a consequence of the classical mechanics.

To calculate the statistical partition function in equation \eqref{eq:Model:Partition} the most important eigenvalues are those with small real part. Therefore we will focus on the narrow bands and treat them in semiclassical approximation.

\subsubsection{Riemann surface and Picard-Fuchs equation}\label{sec:Different:PF}

\begin{figure}
	\includegraphics[width=0.45\textwidth]{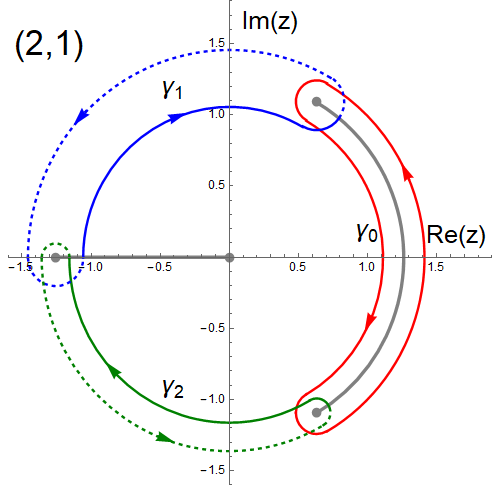}\vspace{0.05\textwidth}
	\includegraphics[width=0.45\textwidth]{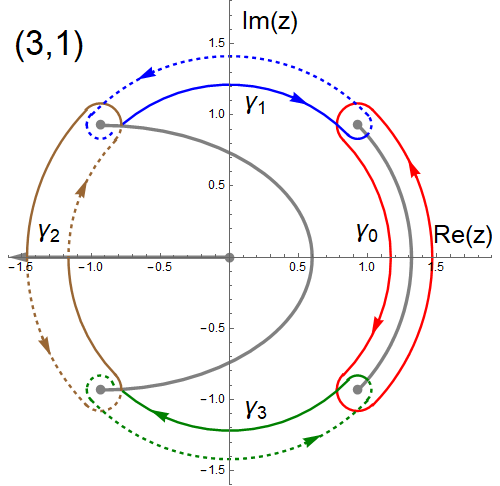}\\
	\includegraphics[width=0.45\textwidth]{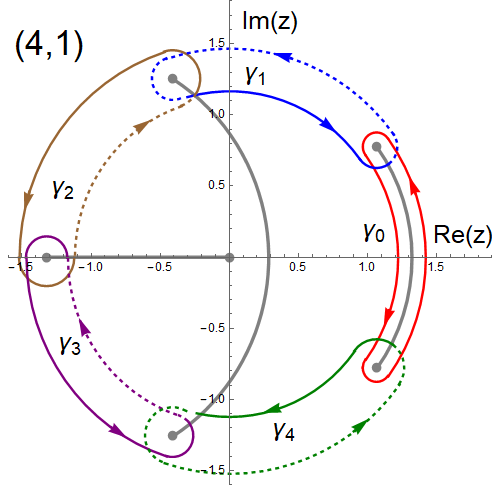}\vspace{0.05\textwidth}
	\includegraphics[width=0.45\textwidth]{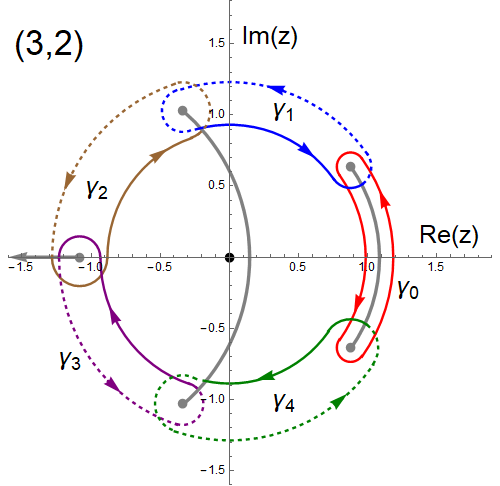}
	\caption{The integration cycles in the complex $z$-plane for the four non-Hermitian cases that are discussed in section \ref{sec:Large:Different}. In all images we set $u=0$. Each color represents one closed cycle of integration. Solid lines denote the sections which lie on the principal sheet, dashed lines the parts on the second sheet. Top left: (2,1); top right: (3,1); bottom left: (4,1); bottom right: (3,2).\\
	Note the differences in the structure of the branch cuts: in the (2,1) case all branch points are finite, while in the (1,1) case in figure \ref{fig:Cycles11} one branch point lies at $\infty$. Similar differences exist between the other three figures, whether the branch points are at finite values of $z$ or at $\infty$, and whether the origin is a branch point or a pole. Reproduced with permission from Ref. \cite{JoPA}.}
	\label{fig:CyclesNonHermitian}
\end{figure}

We use the rescaled energy variable $u$ in equation \eqref{eq:Model:RescaledEnergy}, substitute $z=e^{i\theta}$ in the Hamiltonian \eqref{eq:Model:HamiltonianIso}, and write the classical energy-momentum relation as
\begin{equation}\label{eq:Large:GeneralizedEnergyMomentum}
	u\frac{n_1+n_2}{n_1n_2} = p^2 - \left( \frac{1}{n_1}e^{in_1\theta} + \frac{1}{n_2}e^{-in_2\theta} \right).
\end{equation}
The generalization for the complex algebraic curve in equation \eqref{eq:Large:Riemann11} is the family of curves
\begin{equation}\label{eq:Large:GeneralizedRiemann}
	\mathcal{E}_u:\quad\quad {\cal F}(p,z) = n_1n_2p^2 z^{n_2} - \left(n_2z^{n_1+n_2} + (n_1+n_2)uz^{n_2} + n_1\right) = 0.
\end{equation}
This defines implicitly a double-valued function $p(z)$. It is easy to see that $(\partial {\cal F}/\partial z,\partial {\cal F}/\partial p)$ does not vanish on ${\cal E}_u$ unless $u=-e^{\frac{2\pi im}{n_1+n_2}}$ for an integer $m$. For the non-singular values of $u$ the function $p(z)$ is locally holomorphic except for the points $z=0,\infty,z_j$, where $z_j,j=1,...,n_1+n_2$ are the roots of $p^2=0$. The $z_j$ are the turning points of classical motion in complex coordinates. Near these special points $p(z)$ behaves as
\begin{align}\label{eq:Large:GeneralizedBranchPoints}
	&p\sim z^{-n_2/2},& (z\sim 0)\\\nonumber
	&p\sim z^{n_1/2},& (z\sim \infty)\\\nonumber
	&p\sim (z-z_j)^{1/2}.& (z\sim z_j)
\end{align}
The $z_j$ are $n_1+n_2$ branch points. If $n_2$ ($n_1$) is odd, then $0$ ($\infty$) is an additional branch point; for even $n_2$ ($n_1$) there is a normal pole at $0$ ($\infty$). Hence there are $n_1+n_2+1$ branch points on the Riemann sphere if one of the integers is odd, and $n_1+n_2+2$ branch points if both are odd.\footnote{Here we ignore the case that $n_1,n_2$ are both even, because if both integers can be divided by the same number $n$ we can define $z'=e^{in\theta}$ to obtain a simpler algebraic curve.} In all cases there is an even number of branch points which can be connected pairwise to form branch cuts. For $(n_1,n_2)=(2,1)$ we obtain four branch points and two branch cuts and a Riemann surface of genus 1, as in figure \ref{fig:TorusConstruction}. For $(n_1,n_2)=(3,1),(4,1),(3,2)$ the asymptotic expansions \eqref{eq:Large:GeneralizedBranchPoints} give six branch points. Consequently there are three branch cuts in the complex plane. Through a similar construction as in figure \ref{fig:TorusConstruction} one obtains a Riemann surface which is topologically equivalent to a figure "8", i.e. a figure with two holes and genus 2 \cite{JoPA,thesisPhD}. In the following we consider these four cases because there are no naturally occurring ions with larger charge. However, mathematically the algebraic curves for higher values of the integers can be constructed in the same way, yielding Riemann surfaces with larger genus.

In figure \ref{fig:CyclesNonHermitian} we show the structure of branch points in the $z$-plane for these four cases. On a Riemann surface with genus $g=1(2)$ there are two (four) independent closed cycles \cite{Miranda}. In figure \ref{fig:CyclesNonHermitian} we define three cycles for the $(2,1)$ case, and five cycles for $(4,1)$ and $(3,2)$. This is done for convenience and symmetry reasons. The superfluous cycle can be expressed by the other cycles. For $(2,1)$ the linear combination $\gamma_0-\gamma_1-\gamma_2$ does not contain any of the branch points and is contractible to a point. For $(4,1)$ the trivial cycle is $\gamma_0-\gamma_1+\gamma_2+\gamma_3-\gamma_4 \cong 0$, and for $(3,2)$ we see that $\gamma_0+\gamma_1-\gamma_2-\gamma_3+\gamma_4 \cong 0$. We choose to include the additional cycle because it gives an easy representation for the symmetry relation between the corresponding actions $S_j(u)$, akin to equation \eqref{eq:Large:Symmetry11}. By substituting $z' = e^{-i\phi}z$ and $u' = e^{i\phi}u$ the cycles transform $\gamma_j\to\gamma_{j+1}$. For the $(2,1)$ case the resulting symmetry relation is
\begin{equation}\label{eq:Large:Symmetry21}
	S_0(u) = e^{\pi i/3}S_1(e^{-2\pi i/3}u) = e^{-\pi i/3}S_2(e^{2\pi i/3}u)\,.
\end{equation}
The analogous symmetry relations for the genus-2 cases are shown in Ref. \cite{JoPA}.

To calculate the actions $S(u) = \oint_\gamma\lambda(u)$ we continue in the same manner as in section \ref{sec:Large:Equal}. The 1-form (cf. equation \eqref{eq:Large:1form11}) with general $n_1,n_2$ is
\begin{equation}\label{eq:Large:Generalized1form}
	\lambda(u) = p(\theta)d\theta = p(z)\frac{dz}{iz} = \frac{\left(n_2z^{n_1+n_2} + (n_1+n_2)uz^{n_2} + n_1\right)^{1/2}} {i\sqrt{n_1n_2}z^{1+n_2/2}} dz.
\end{equation}
On a Riemann surface of genus $g=1(2)$ there are two (four) independent closed cycles. According to the de Rham theorem, this is equal to the number of linearly independent 1-forms, modulo exact forms. Therefore a set of the 1-form \eqref{eq:Large:Generalized1form} and its first few derivatives, $\{\partial_u^k \lambda(u)\}_{k=0}^K$, is linearly dependent if it contains the first $K=2(4)$ derivatives. We build a linear combination of these which equals an exact form (for details see \cite{JoPA}). The integral of the exact form along a closed cycle gives zero. What is left is a linear combination of the action and its first derivatives, cf. equation \eqref{eq:Large:PF11}. In the $(2,1)$ case we find this Picard-Fuchs equation as
\begin{equation}\label{eq:Large:PF21}
	(u^3+1)S_j''(u)+\frac{u}{4}S_j(u)=0.
\end{equation}
This is a second-order differential equation. The Picard-Fuchs equations for the genus-2 cases are fourth-order ODEs which can be found in Ref. \cite{JoPA}. Equation \eqref{eq:Large:PF21} admits two solutions $F_0(u^3)$ and $uF_1(u^3)$ which are given in terms of the hypergeometric functions \cite{JoPA,Mathworld}
\begin{eqnarray}\label{eq:Large:Solutions21}
	F_0(u^3) &=& _2F_1\left(-\frac{1}{6},-\frac{1}{6};\frac{2}{3};\,-u^3\right),
	\\\nonumber
	F_1(u^3) &=& _2F_1\left(+\frac{1}{6},+\frac{1}{6};\frac{4}{3};\,-u^3\right).
\end{eqnarray}
The actions are formed from a linear combination of these, $S_j(u)=C_{j0}F_0(u^3)+C_{j1}uF_1(u^3)$. Expanding the hypergeometric functions near the origin, $F_{0,1}(u^3)=1+\mathcal{O}(u^3)$, one notices that $S_j(u)=C_{j0}+uC_{j1}+\mathcal{O}(u^3)$ as $u\to 0$. The constants $C_{0k}$ are therefore given by $C_{00}=S_0(0)$ and $C_{01}=S_0'(0)$. Straightforward integration and the symmetry relation \eqref{eq:Large:Symmetry21} yield
\begin{eqnarray}\label{eq:Large:Coefficients21}
	C_{00} = C_{10}e^{\pi i/3} = C_{20}e^{-\pi i/3}
	&=& \frac{2^{11/6}3\pi^{3/2}}{\Gamma(\frac{1}{6})\Gamma(\frac{1}{3})},
	\\\nonumber
	C_{01} = C_{11}e^{-\pi i/3} = C_{21}e^{\pi i/3}
	&=& \frac{3^{1/2}\Gamma(\frac{1}{6})\Gamma(\frac{1}{3})}{2^{11/6}\pi^{1/2}}.
\end{eqnarray}
The actions $S_j(u)$ for $(n_1,n_2) = (2,1)$ are fully given by equations \eqref{eq:Large:Symmetry21}, \eqref{eq:Large:Solutions21}, and \eqref{eq:Large:Coefficients21}. The analogous expressions for the genus-2 cases with $(n_1,n_2)=(3,1),(4,1),(3,2)$ are given in Ref. \cite{JoPA}. In the next section we discuss how to obtain semiclassical results for the physical observables.

\subsubsection{Semiclassical results in the non-Hermitian cases}\label{sec:Different:Semiclassical}
In this section we calculate the eigenenergies and bandwidths of the non-Hermitian Hamiltonian in equation \eqref{eq:Model:HamiltonianIso} with the Bohr-Sommerfeld quantization condition and Gamow's formula. To utilize these standard semiclassical results we need to calculate the classical and the instanton actions, $S_{cl,inst}(u) = \oint_{\gamma_{cl,inst}}\lambda(u)$. The crucial part hereby is identifying the correct cycle of integration. In section \ref{sec:Large:Equal}, when discussing the case of a Hermitian Hamiltonian, we identified these with trajectories which connect the classical turning points through the classically allowed or forbidden region respectively, cf. figure \ref{fig:Cycles11}. In the non-Hermitian case this is not so clear, because there exist more than two turning points, and in the space with complex coordinate, momentum, and energy the concept of classically allowed or forbidden region doesn't apply. Instead, to identify the correct actions $S_{cl,inst}(u)$ we look at the analytic behavior of these actions near special values of the energy $u$.

\begin{figure}
	\includegraphics[width=0.48\textwidth]{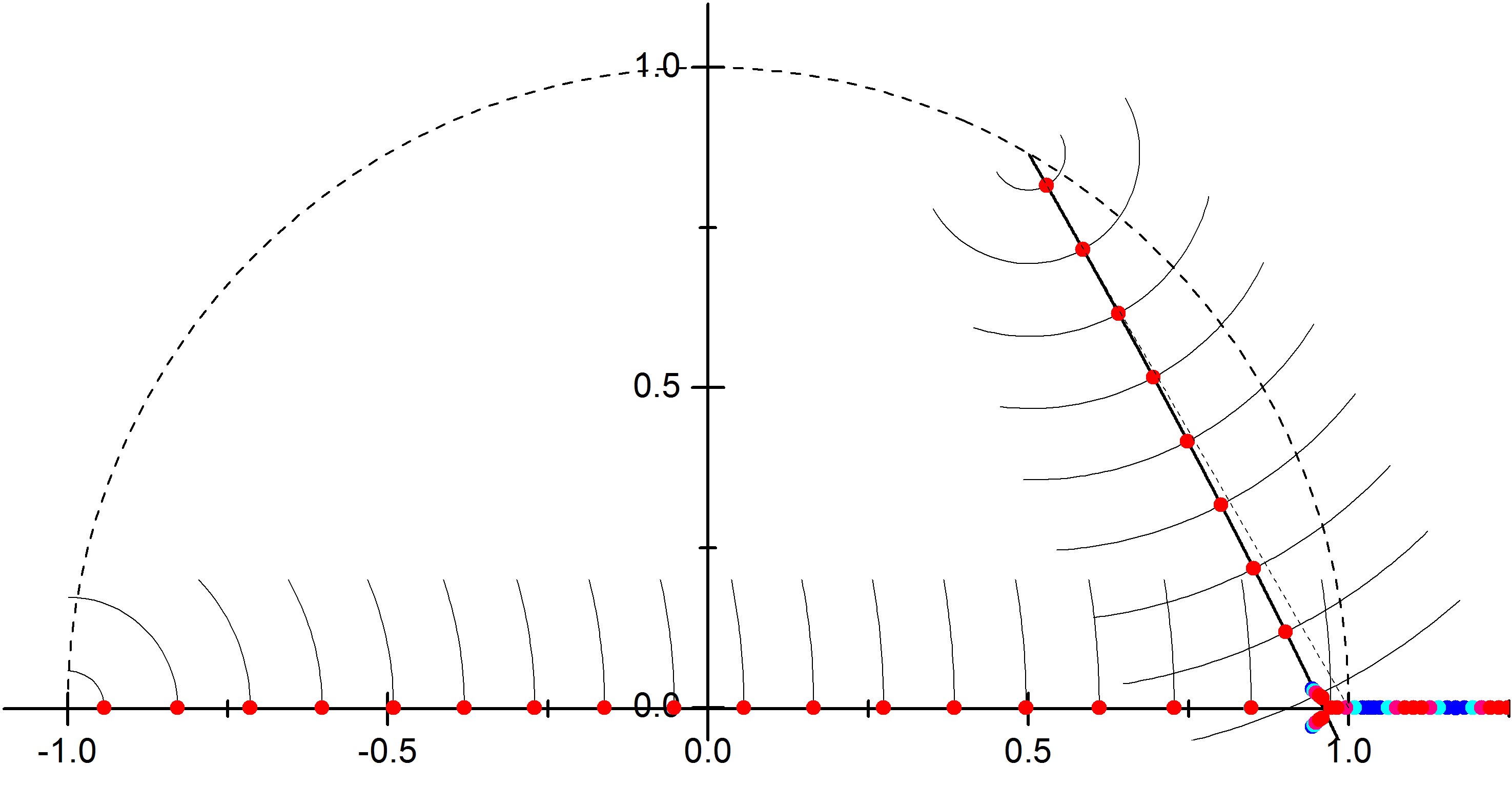}\hspace{0.03\textwidth}
	\includegraphics[width=0.48\textwidth]{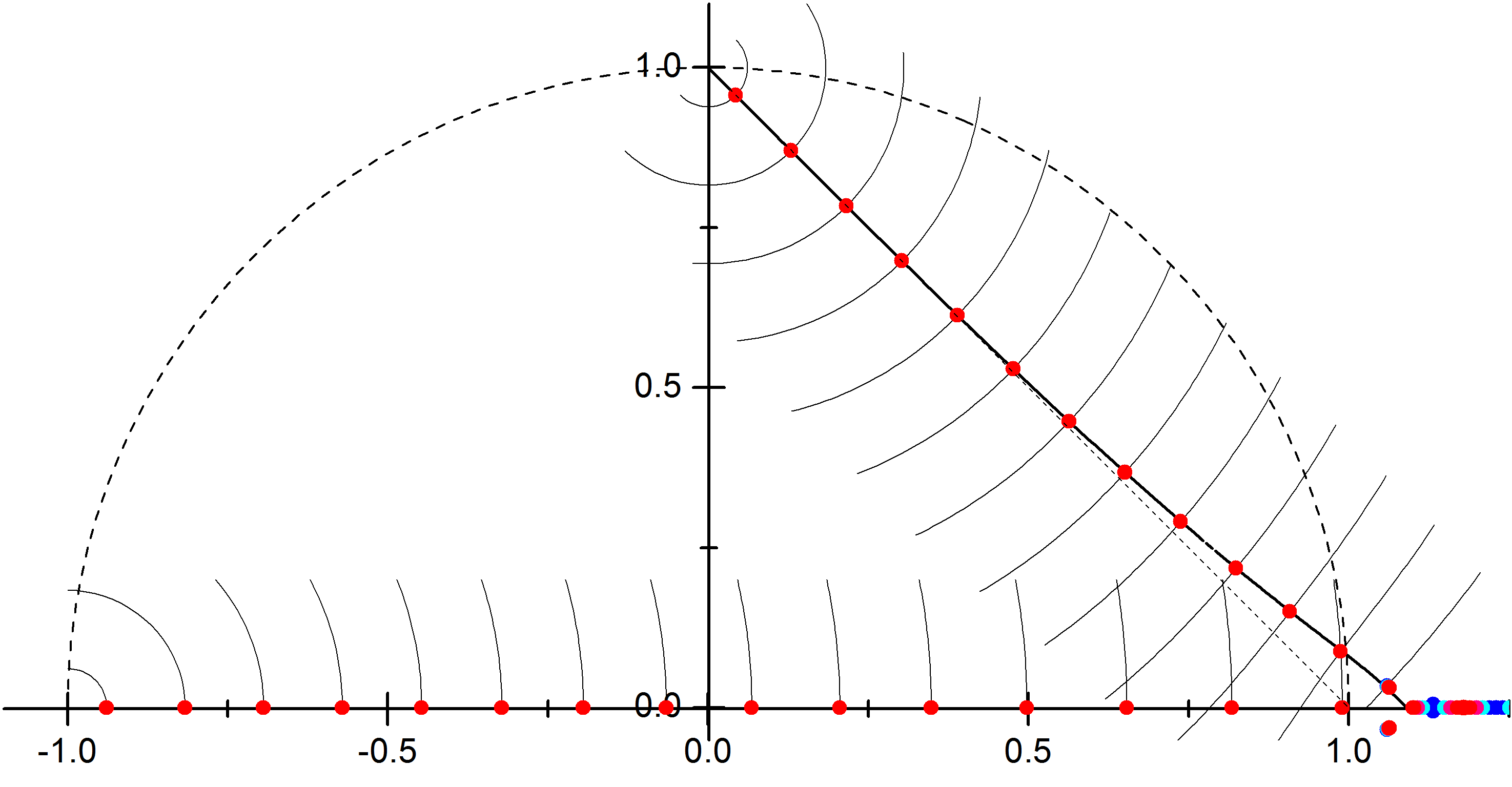}\\
	\includegraphics[width=0.48\textwidth]{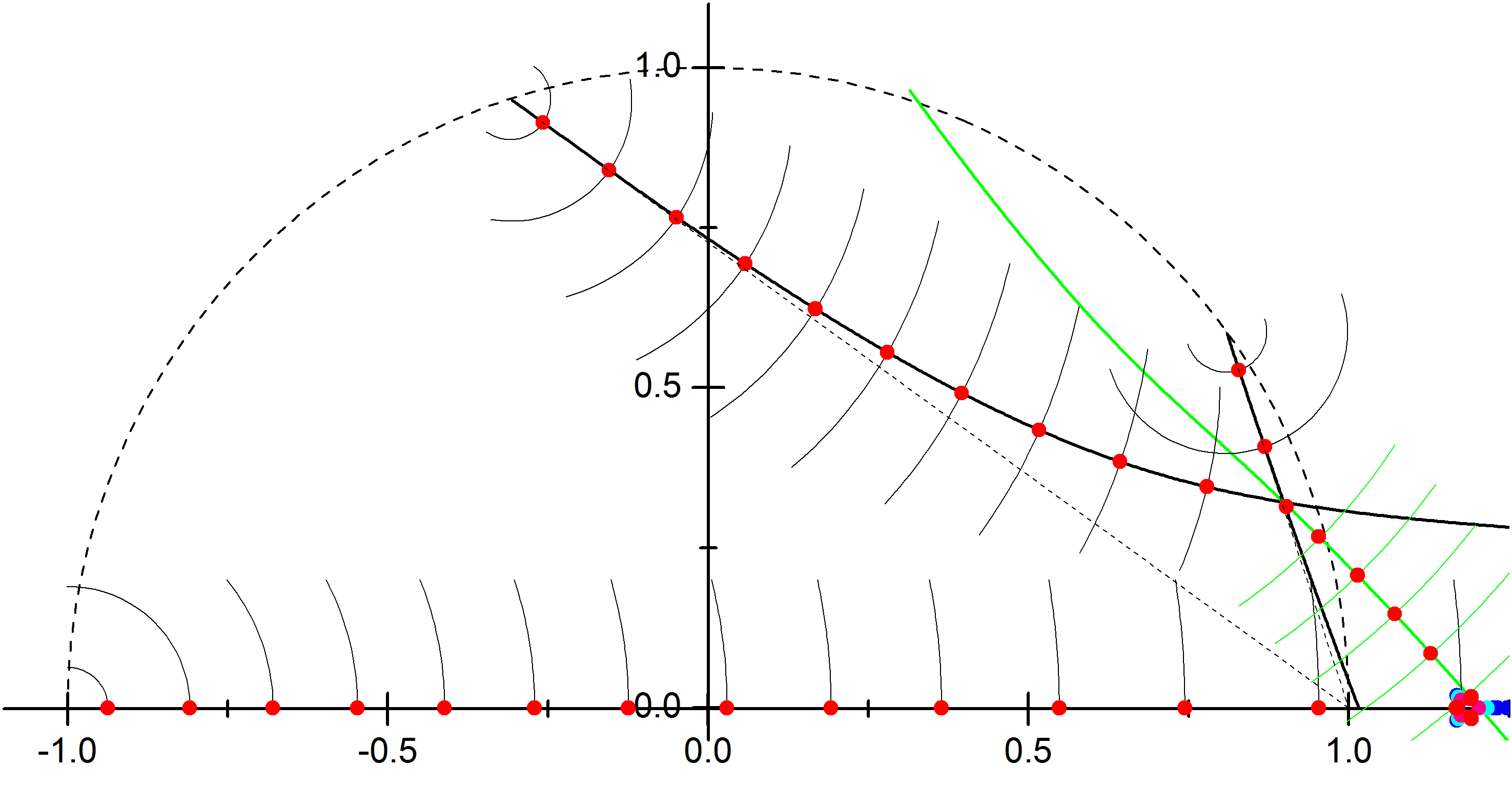}\hspace{0.03\textwidth}
	\includegraphics[width=0.48\textwidth]{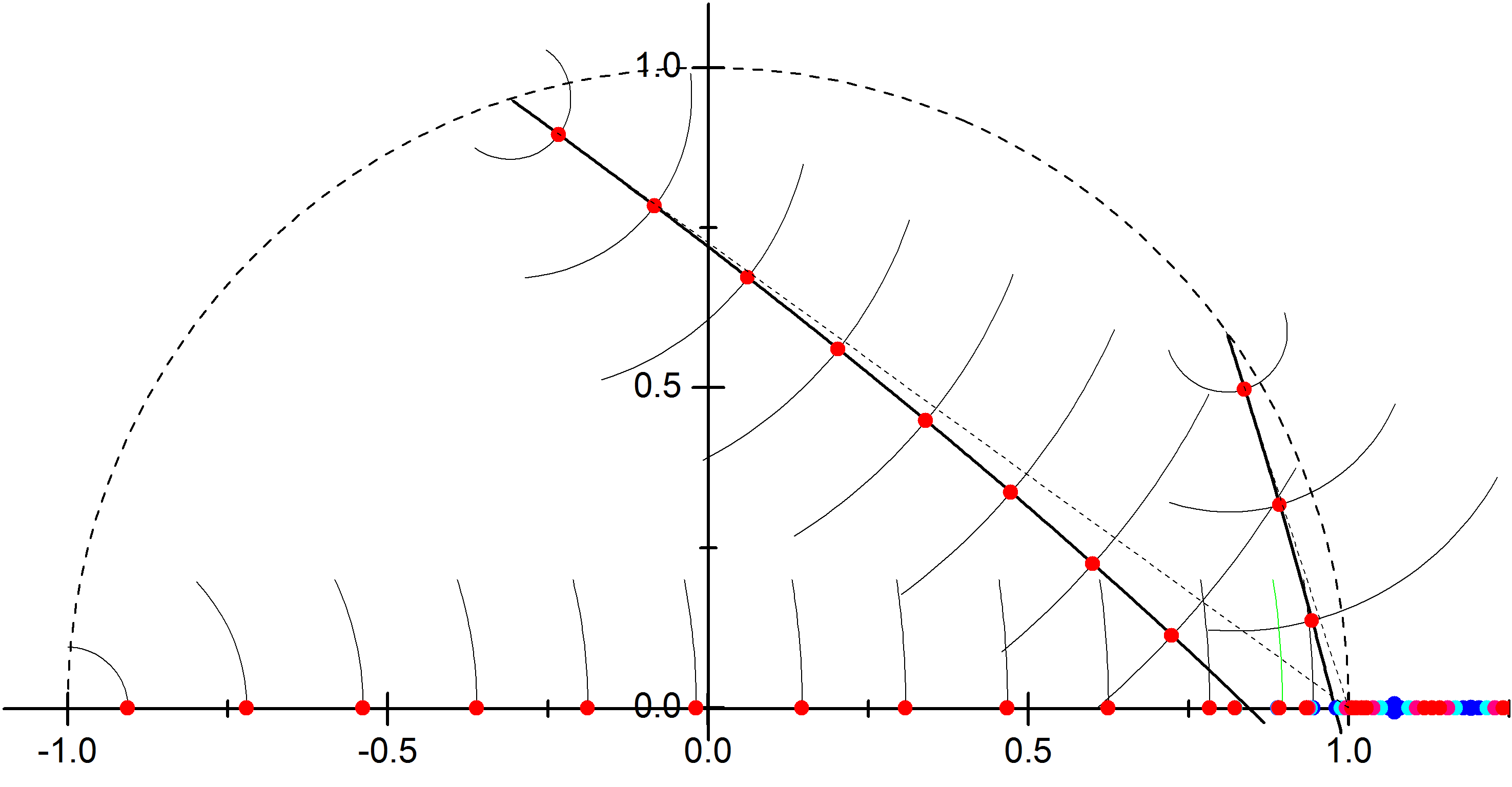}
	\caption{Narrow energy bands (red dots) in the upper half-plane of complex energy $u$ for large $\alpha$, cf. figure \ref{fig:Spectrum}. In all four cases, $Im\, S_0(u)=0$ along the real axis, where the thin lines mark $|S_0(u)|=2\pi\alpha^{-1/2}(m+1/2)$, the quantization condition. The other black lines mark $Im\, S_j(u)=0$ for the other actions $S_j(u)$, and the thin perpendicular lines mark $|S_j(u)|=2\pi \alpha^{-1/2}(m+1/2)$. In all cases $S_j(u)$ corresponds to an action encircling two branch points. These points coalesce at a singular value of $u$ on the unit circle (dashed) where the spectral branch ends. Near intersections of two lines neither quantization condition holds, cf. $u\approx0.90+0.31i$ in $(4,1)$ and $u\approx 0.82$ in $(3,2)$. Beyond this intersection the states are quantized according to the sum of the two corresponding actions, $S_1+S_2$ in $(4,1)$ and $S_2+S_3$ in $(3,2)$, marked in green. To the right all lines coalesce and beyond this point we observe wide bands with narrow gaps. The lower half-plane shows the mirror image (i.e. complex conjugate) of the upper half plane.\\
	Top left: $(n_1,n_2)=(2,1)$, $\alpha=200$; top right: $(3,1)$, $\alpha=300$; bottom left: $(4,1)$, $\alpha=400$; bottom right: $(3,2)$, $\alpha=400$. Reproduced with permission from Ref. \cite{JoPA}.}
	\label{fig:Quantization}
\end{figure}

The Bohr-Sommerfeld condition requires that the classical action goes to zero at the classical minimum of the potential. This happens when two turning points collide which causes the corresponding cycle of integration to collapse to a point. We can easily check that in all four cases in figure \ref{fig:CyclesNonHermitian} the cycle $\gamma_0$ collapses to a point as $u\to-1$. The corresponding action goes to zero, $S_0(-1)=0$. Therefore we identify $S_0(u)$ as the classical action which quantizes into the branch of eigenstates that terminates at $u=-1$. It follows immediately from the symmetry relation \eqref{eq:Large:Symmetry21} that at the singular point $u=e^{i\pi/3}$ ($e^{-i\pi/3}$) the cycle $\gamma_1$ ($\gamma_2$) collapses to a point and the action $S_{1}(u)$ ($S_2(u)$) goes to zero for $(n_1,n_2)=(2,1)$. It should be thus identified with the classical action for the spectral branch terminating at $u=e^{i\pi/3}$ ($e^{-i\pi/3}$). In the same manner the analogous symmetry relations for the genus-2 cases in Ref. \cite{JoPA} allow to identify the classical actions for all the spectral branches in figure \ref{fig:Spectrum}. Quantizing the classical actions according to the Bohr-Sommerfeld rule,
\begin{equation}\label{eq:Large:BSNonHermitian}
	S_j(u_m^{(j)})=2\pi\alpha^{-1/2}(m+1/2), \quad \quad m=0,1,...\, ,
\end{equation}
one finds the semiclassical energies $u_m^{(j)}$ determining the $q=0$ edges of the narrow bands in the complex plane. The agreement with the numerical data is visualized in figure \ref{fig:Quantization}. The excellent agreement holds all the way up to the point where all spectral branches coalesce. Beyond this point the semiclassical approximation breaks down, which manifests in e.g. the appearance of wide Bloch bands.

All graphs exhibit spectral branches along the lines where one of the actions $S_j(u)$ is real, while the narrow bands lie at the points determined by the Bohr-Sommerfeld condition \eqref{eq:Large:BSNonHermitian}. For $(2,1)$ and $(3,1)$ there exists a total of three spectral sequences, for $(4,1)$ and $(3,2)$ five sequences due to a higher number of special energies. In the $(4,1)$ case the two complex-valued branches intersect at $u\approx0.90+0.32i$. Beyond this point the two sequences merge into one, for which the quantization condition is neither determined by $S_1$ nor $S_2$ individually, but instead by the sum $S_1+S_2$ (shown in green). For $(3,2)$ the two lines for the complex-conjugate pair $S_2$ and $S_3$ collide at $u\approx0.84$, the other pair collides at $u\approx0.98$ where the semiclassical approximation breaks down. A closer look at the state at $u\approx0.89$ reveals that this cannot be explained by the quantization of $S_0$ along the real axis. However, it meets the Bohr-Sommerfeld condition \eqref{eq:Large:BSNonHermitian} for $S_2+S_3$ with $m=17$. Thus we may conclude that the spectral branches can be derived from the Bohr-Sommerfeld condition for one of the actions, or upon intersection of two branches by the sum of the two actions of these branches.

To calculate the width of these bands with Gamow's formula, 
\begin{equation}\label{eq:Large:GamowGeneral}
	(\Delta u)_m=\frac{\omega}{\pi\sqrt{\alpha}}\, e^{i\alpha^{1/2} S_{inst}(u_m)/2}\,,
\end{equation}
we need to identify the instanton actions.\footnote{The classical frequency $\omega$ is determined from the harmonic oscillator approximation, i.e. by expanding the potential around $\theta=0$.} In Hermitian quantum mechanics the instanton trajectory connects the two classical turning points through the classically forbidden region, cf. figure \ref{fig:Cycles11}. Hence we identify the instanton cycle as the other possible cycle that connects the same two turning points. This is a combination of all other integration cycles. $\gamma_i$ The instanton actions that correspond to the classical actions $S_0(u)$ are
\begin{eqnarray}\label{eq:Large:InstantonNonHermitian}
	S_{inst} (u) =& - S_1(u) + S_2(u), & (2,1);\\\nonumber
	S_{inst} (u) =& - S_1(u) - S_2(u) + S_3(u), & (3,1);\\\nonumber
	S_{inst} (u) =& - S_1(u) - S_2(u) + S_3(u) + S_4(u), & (4,1);\\\nonumber
	S_{inst} (u) =& - S_1(u) + S_2(u) - S_3(u) + S_4(u), & (3,2).
\end{eqnarray}
From the symmetry relation \eqref{eq:Large:Symmetry21} between the actions and its analogons for the genus-2 cases it is easy to check that these combinations are purely imaginary, which makes the bandwidth in equation \eqref{eq:Large:GamowGeneral} real, as required.

\begin{figure}
	\includegraphics[width=0.9\textwidth]{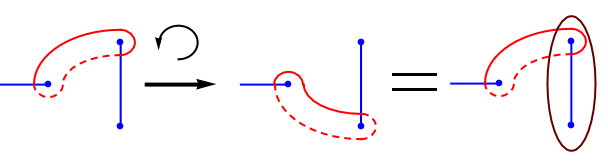}
	\caption{In a monodromy transformation the parameter $u$ is smoothly changed around a critical value in parameter space and returned to its original value, e.g. $(1+u) \to (1+u)e^{2\pi i}$. During the transformation the branch points (blue) move in the complex plane, and the same structure of branch points is recovered. However, if a special value of the parameter $u$ is enclosed by the trajectory in parameter space, e.g. $u=-1$, then the two branch points which collide at $u=-1$ are exchanged. During the transformation the integration cycle (red) is not allowed to cross a branch point, hence they are pulled along with the branch points. To restore the original cycle a closed cycle enclosing the two branch points has to be added.}
	\label{fig:Monodromy}
\end{figure}

More can be said when considering the analytic structure of the classical and instanton action in a vicinity of $u=-1$. Therefore we use a concept called monodromy \cite{Miranda,EncyclopediaMath}, which is visualized in figure \ref{fig:Monodromy}. We choose some $u\gtrsim -1$ and allow $u$ to wind around $-1$ (i.e. $(u+1)\to(u+1)e^{2\pi i}$). The two branch points inside the cycle $\gamma_0$ in figure \ref{fig:CyclesNonHermitian} are exchanged by this transformation via a counter-clockwise half-turn; the branch cut in effect rotates by $180^\circ$. For $\gamma_0$ this has no effect, the cut turns within it. Not so for $\gamma_1$: if this cycle is never to intersect the branch points, it is continuously deformed and as a result of this monodromy transformation we obtain $\gamma_1\to\gamma_1+\gamma_0$, thus $S_1$ picks up a contribution of $S_0$. This effect is visualized in figure \ref{fig:Monodromy}. While we have returned to the initial value of $u$, the period $S_1$ does not return to its original value and thus can't be analytic. This occurs for every monodromy cycle near $u=-1$. The only function which monotonically increases as the phase of its argument grows is the complex logarithm. Thus $S_1$ must have a logarithmic dependence on $1+u$. One can check that
\begin{equation}\label{eq:Large:Monodromy}
	S_1(u) = Q_1(u) - \frac{i}{2\pi}S_0(u)\ln(1+u)
\end{equation}
yields the correct behavior, where $Q_1(u)$ and $S_0(u)$ are analytic functions of $(1+u)$. The same applies to the other cycle which is connected to the same branch cut. Therefore the instanton action $S_{inst}$ in equation \eqref{eq:Large:InstantonNonHermitian} picks up a contribution of $-2S_0$. Hence we can derive the Bohr-Sommerfeld quantization condition \eqref{eq:Large:BSNonHermitian} from the requirement that the monodromy transformation leaves the bandwidth \eqref{eq:Large:GamowGeneral} unchanged.

\begin{figure}
	\centering
	\includegraphics[width=0.6\textwidth]{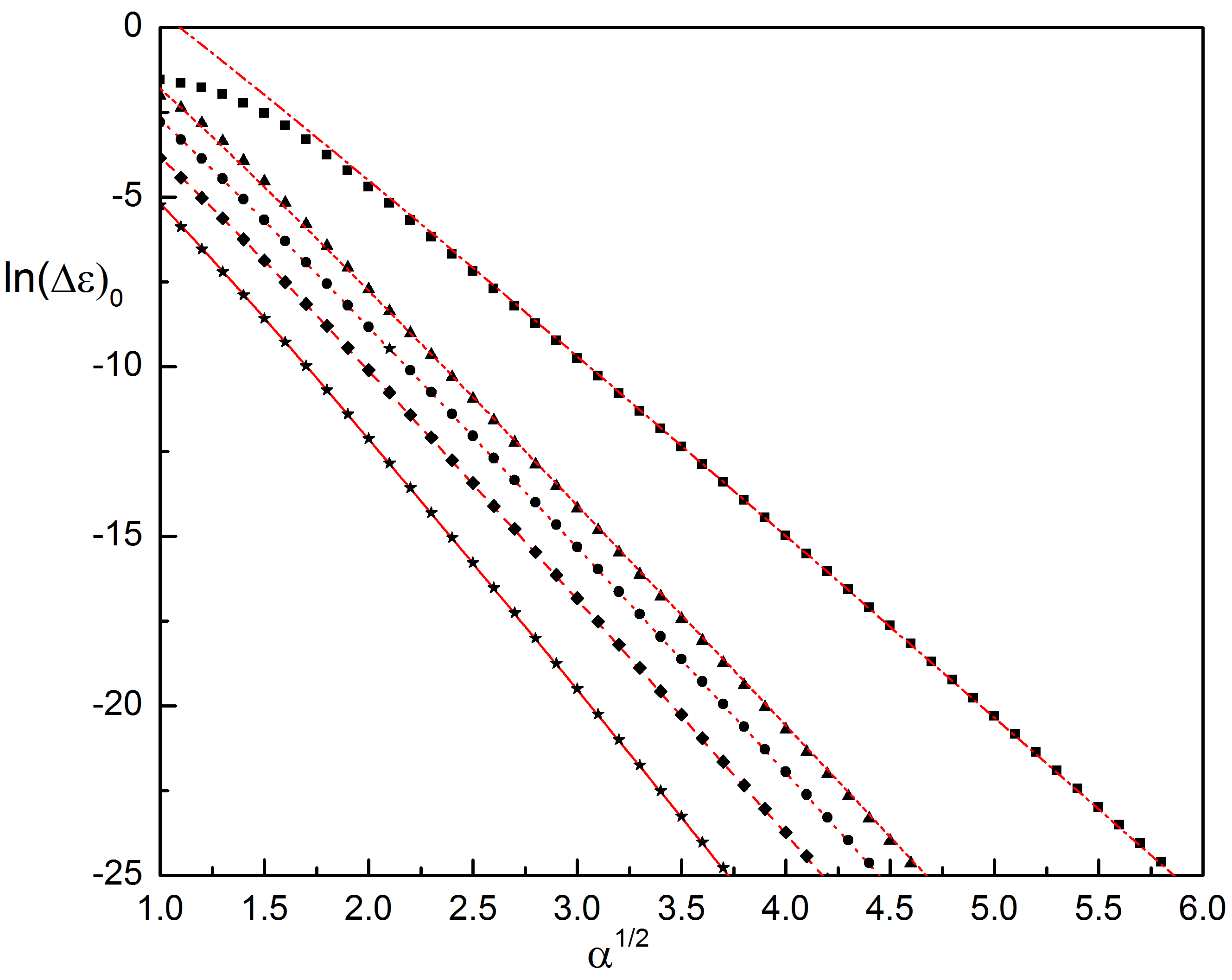}
	\caption{Analytic (numerical) results for the logarithm of the bandwidth of the lowest band, $\ln(\Delta\varepsilon)_0$, as a function of $\alpha^{1/2}$, for all five cases with Riemann surfaces of genus 1 or 2. $(1,1)$: solid line (stars), $(2,1)$: dashed line (diamonds), $(3,1)$: dotted line (circles), $(4,1)$: short-dashed line (triangles), and $(3,2)$: dash-dotted line (squares). Reproduced with permission from Ref. \cite{JoPA}.}
	\label{fig:Bandwidth}
\end{figure}

A comparison of the results for the bandwidth with numerical simulations is shown in figure \ref{fig:Bandwidth} for the four non-Hermitian cases and the Hermitian $(1,1)$ case. All cases show good agreement with the numerical data already for moderate values of the parameter $\alpha$.\footnote{Note however, that for the genus-2 cases Gamow's formula had to be multiplied by an overall factor of $3/2$ (in $(3,1)$ case) or $2$ (in $(4,1)$ and $(3,2)$ cases), respectively. The origin of this preexponential factor is beyond the scope of this paper.}

To summarize, we find that in all cases the bandwidth is of the form
\begin{equation}\label{eq:Large:GamowForm}
	(\Delta \varepsilon)_m = A\times\left(\frac{k}{m+1/2}\right)^{(m+1/2)}\times \exp\left(-b\sqrt{\alpha}+(m/2+3/4)\ln\alpha\right).
\end{equation}
The pressure, which is calculated from the lowest eigenvalue, contains the ideal gas pressure and the Debye-Hueckel correction,
\begin{equation}\label{eq:Large:PressureForm}
	P = C k_BTf - c \sqrt{k_BTeE_0f}.
\end{equation}
Here $A$, $k$ and $b$, and $C$ and $c$, are numerical factors that can be calculated directly by expanding $S_0$ and $S_{inst}$:\vspace{0.5ex}\\
\begin{tabular}{c||c|c|c|c|c}
	$(n_1,n_2)$ & $A$ & $k$ & $b$ & $C$ & $c$\\\hline\hline
	$(1,1)$ & $4/\pi$ & $32e$ & $8$ & $2$ & $1$ \\\hline
	$(2,1)$ & $2\sqrt{6}/\pi$ & $36\sqrt{6}e$ & $3\sqrt{6}$ & $3/2$ & $\sqrt{3/2}$ \\\hline
	$(3,1)$ & $4\sqrt{2}/\pi$ & $6.35$ & $7.06$ & $4/3$ & $\sqrt{2}$ \\\hline
	$(4,1)$ & $5\sqrt{5/2}/2\pi$ & $1303.46$ & $6.90$ & $5/4$ & $\sqrt{5/2}$ \\\hline
	$(3,2)$ & $5\sqrt{5/2}/3\pi$ & $6740.06$ & $5.65$ & $5/6$ & $\sqrt{5/2}$ \\
\end{tabular}\vspace{1ex}\\
These values quantify the thermodynamic properties of the ion channels for all five different combinations of charged ions which give a Riemann surface of genus 1 or 2. With a maximum valency of 4 these are also the physically relevant cases. Most importantly we show that the Coulomb gas with unequal valency $n_1 \neq n_2$ has the same qualitative behavior as the standard gas with ions of equal valency, $n_1=n_2$. In all cases the pressure consists of the ideal gas pressure and the Debye-Hueckel correction, see equation \eqref{eq:Large:PressureForm}. Crucially for the biological functions of the ion channel, in all cases the bandwidth shows exponential decay with the square-root of the fugacity $\alpha$ and has a universal pre-exponential factor of $\alpha^{3/4}$. However the factor in the exponent shrinks when the valency is increased, meaning that the transport barrier falls off slower when transporting ions with larger valency.

\subsection{Higher-order corrections from exact WKB method}\label{sec:Large:Secondorder}
The approximations for the eigenvalues of the non-Hermitian Hamiltonian can be improved further by considering second- and higher-order terms in the WKB series. The inspiration comes from the exact WKB method which was studied extensively in the context of resurgence theory \cite{BenderOrszag,BasarDunneUnsal}. We use this to get a better approximation for the eigenvalues, and with that the pressure of the Coulomb gas, at moderate values of the charge concentration $\alpha\gtrsim 1$. The key is that the $q=0$ band edge, which gives the pressure in equilibrium, is determined by an infinite series in $\alpha^{-1}$ (i.e. $\hbar^2$ in usual quantum mechanics),
\begin{equation}\label{eq:Large:ExactWKB}
	\sum_{n=0}^\infty \frac{(-1)^n}{\alpha^n} \oint_{\gamma_{cl}} \rho_{2n}(\theta,u_m) d\theta = \frac{2\pi(m+1/2)}{\sqrt\alpha}.
\end{equation}
$\rho_0(\theta,u) = p(\theta,u)$ is the classical momentum, and the other terms can be found through a recursive relation \cite{BenderOrszag}. Equation \eqref{eq:Large:ExactWKB} is sometimes also referred to as the generalized Bohr-Sommerfeld quantization condition. In 2017 \cite{BasarDunneUnsal} calculated the exact WKB series at all orders for a class of genus-1 cases which include the cosine potential, i.e. the Hermitian $(1,1)$ case in our notation. Here we follow the ideas in \cite{KreshchukGulden} and chapter 5 of \cite{thesisPhD} which give a general procedure to calculate the terms order-by-order for any potential, and can also be applied to non-Hermitian Hamiltonians.

It is evident that truncation of equation \eqref{eq:Large:ExactWKB} at the $n=0$ term leads to the usual Bohr-Sommerfeld quantization condition. To improve we include the $n=1$ term. The integrand is given by
\begin{equation}\label{eq:Large:SecondOrder}
	\rho_2(\theta,u) d\theta =
	\left( \frac{\partial_\theta^2(\rho_0(\theta,u)^2)}{48\rho_0(\theta,u)^3} 
	+ \frac{5}{24} \partial_\theta \frac{\rho_0'(\theta,u)}{\rho_0(\theta,u)^2} \right)d\theta,
\end{equation}
where the prime denotes a derivative with respect to $\theta$ \cite{BenderOrszag}. The second term is an exact form which integrates to zero. We drop this exact form, use the expression \eqref{eq:Large:GeneralizedEnergyMomentum} for the classical momentum $p=\rho_0$, and perform the coordinate transformation $z=e^{i\theta}$ to write the second-order 1-form as
\begin{equation}\label{eq:Large:SecondOrderz}
	\tilde{\rho}_2(z,u) dz = \frac{-n_1z^{n_1}-n_2z^{-n_2}} {48\left( u\frac{n_1+n_2}{n_1n_2} + \frac{1}{n_1}z^{n_1} + \frac{1}{n_2}z^{-n_2}\right)^{3/2}iz} dz.
\end{equation}
A comparison with equation \eqref{eq:Large:Generalized1form} shows that the second-order 1-form $\tilde{\rho}_2(z,u)dz$ has the same branch points as the action 1-form $\lambda(u)$. Therefore it is defined on the same Riemann surface. As discussed in the preceding sections, on the Riemann surfaces of genus $g = 1 (2)$ there exist two (four) linearly independent 1-forms, up to an exact form. We take $\{\partial_u^k \lambda(u)\}_{k=0}^K$ as this maximal independent set with $K = 1 (3)$. This forms a basis for the space of all 1-forms. Hence the second-order correction can be written as a linear combination of these basis 1-forms, modulo an exact form. We find this linear combination in the same way as in the derivation of the Picard-Fuchs equations \eqref{eq:Large:PF11} and \eqref{eq:Large:PF21} and integrate it along the classical cycle $\gamma_{cl}$ to get
\begin{equation}\label{eq:Large:SecondOrderResult}
	\oint_{\gamma_{cl}} \tilde{\rho}_2(z,u) dz = -a \left( S_0'(u) + 2u S_0''(u) \right), \quad
	\begin{tabular}{c||c|c|c|c|c}
		$(n_1,n_2)$ & (1,1)  & (2,1)  & (3,1)  & (4,1)  & (3,1)\\\hline
		$a$               & 1/48  & 1/18  & 3/32  & 2/15  & 3/10
	\end{tabular}.
\end{equation}

These expressions fully define the second-order corrections in terms of the classical action and its derivatives with respect to $u$. These are easily obtained from the previous results, equations \eqref{eq:Large:Solutions11}-\eqref{eq:Large:Coefficients11} and \eqref{eq:Large:Solutions21}-\eqref{eq:Large:Coefficients21} (see Ref. \cite{JoPA} for the genus-2 cases). Note that in the genus-1 cases the second derivative $S_0''(u)$ can be replaced with $S_0(u)$ by using the Picard-Fuchs equations \eqref{eq:Large:PF11} and \eqref{eq:Large:PF21}.

\begin{figure}
	\includegraphics[width=0.8\textwidth]{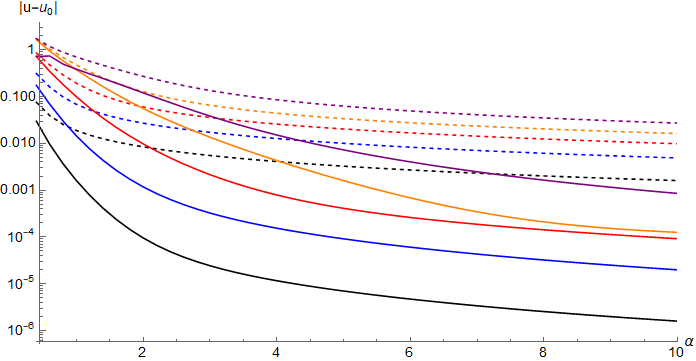}
	\caption{Deviation of the first-order (dashed line) and second-order (solid line) WKB result from the exact numerical result for the lowest eigenvalue as a function of $\alpha$. We show the five different cases: (1,1) in black, (2,1) in blue, (3,1) in red, (4,1) in orange, (3,2) in purple. The error drops by several orders of magnitude when taking the second-order WKB term into account. The approximations converge to the exact result as $\alpha\to\infty$, however already at moderate values of $\alpha\gtrsim1$ the approximations give quite accurate results.}
	\label{fig:SecondOrder}
\end{figure}

Here we want to stress that calculation of the second-order (and any higher) correction is only as computationally demanding as deriving the Picard-Fuchs equation. It does not require solving the differential equation and matching boundary conditions because the correct classical action was already identified. Therefore this can also be used as a simple method to simply calculate the higher-order WKB terms if the classical action was obtained in a different manner. The improvement in the approximation of the lowest eigenvalue is shown in figure \ref{fig:SecondOrder}.

With the second-order result we can calculate the eigenvalues $u$ up to order $\alpha^{-1}$. Therefore we expand the classical action $S_0(u)$ for $u\gtrsim-1$ to order $(u+1)^2$ and solve for $u$. Taking the lowest eigenvalue $u_0$ and applying this to the formula for the pressure \eqref{eq:Model:Pressure} gives
\begin{equation}
P = c_0 k_BTf - c_1 \sqrt{eE_0k_BTf} - c_2 eE_0,
\end{equation}
with the following constants:\\
\begin{tabular}{c||c|c|c}
	$(n_1,n_2)$ & $c_0$ & $c_1$ & $c_2$ \\\hline\hline
	$(1,1)$ & $2$ & $1$ & 1/16\\\hline
	$(2,1)$ & $3/2$ & $\sqrt{3/2}$ & 1/9\\\hline
	$(3,1)$ & $4/3$ & $\sqrt{2}$ & 19/144\\\hline
	$(4,1)$ &  $5/4$ & $\sqrt{5/2}$ & 1/8 \\\hline
	$(3,2)$ & $5/6$ & $\sqrt{5/2}$ & 13/36\\
\end{tabular}\vspace{0.5ex}\\
This gives the ideal gas pressure and the Debye-Hueckel correction from the usual Bohr-Sommerfeld condition. The second-order WKB term gives an additional correction which is independent of the fugacity but only depends on the geometric properties of the channel which are included in the definition of $E_0$.

\section{Summary of semiclassical results}\label{sec:Summary}

In this review we discussed analytic calculations of the thermodynamic properties of an ion channel at large charge concentrations, with an extension to moderate concentrations. We started with discussing a standard mapping of a statistical system onto an effective quantum mechanics \cite{EdwardsLenard,AltlandSimons}. When performing this mapping there is no guarantee that the resulting effective Hamiltonian is Hermitian. Physically one needs to obtain a real and positive partition function. This is e.g. guaranteed if the Hamiltonian obeys $\mathcal{PT}$-symmetry and its lowest eigenvalue is purely real.

Translation between the quantum results and thermodynamic observables is straightforward. Most importantly, the pressure (i.e. free energy density) is given by the quantum mechanical ground-state energy. The adiabatic transport barrier is the width of the lowest Bloch band. The excited states complex energies, c.f. figures \ref{fig:Spectrum} and \ref{fig:Quantization}, describe  higher-order correlation functions. Their imaginary part is responsible for spatial oscillations, while the real part yields the overall exponential decay. Such decaying oscillatory correlation functions reflect short-range charge density wave ionic order within the channel. As seen in figures \ref{fig:Spectrum} and \ref{fig:Quantization}, the onset of complex eigenvalues happens at lower energies for larger valencies of the ions, which implies stronger charge density fluctuations. In all cases we observe that an increase of the charge concentration leads to an exponential reduction of the transport barrier, however this decay is slower if the ion valencies are large. This is visualized in figure \ref{fig:Bandwidth}.

The approximation with the effective 1D Coulomb potential, equation \eqref{eq:Intro:potential}, works best at large ion concentration. Electric field lines leak out of the channel after a characteristic length $\xi$ which is given by $\xi^2=a^2\kappa_1/(2\kappa_2)\ln(2\xi/a)$, where $a$ is the radius of the channel and $\kappa_1,\kappa_2$ are the dielectric constants of water and the surrounding medium. Therefore the 1D Coulomb potential best approximates the situation where the characteristic distance between the ions is small. This is the case of large charge concentration, which is also the case when then semiclassical approximation is applicable.

Here we discussed a method how to perform semiclassical calculations without the need to solve the classical equations of motion and without direct integration. This is particularly useful in the non-Hermitian cases when the solutions to the equations of motions are hardly attainable. Instead we derived and solved the Picard-Fuchs differential equation, which is a tool from algebraic topology. The power of the Picard-Fuchs equation is that it is a coordinate-free expression, i.e. one does not need to know the classical trajectories. In the last part we extended our calculations to second- and higher-order terms in the WKB series. These provide a clearly improved approximation for the eigenvalues especially at moderate charge concentrations, see figure \ref{fig:SecondOrder}.

The applicability of the Picard-Fuchs method extends far beyond the case of ion channels. It can be a powerful tool for Hermitian and non-Hermitian systems alike. Especially the extension to second- and higher-order terms in the WKB series requires very little computational effort once the classical action has been calculated. We believe that this method can be especially useful for many non-Hermitian systems that appear after mapping statistical mechanics onto an effective quantum theory.

\vspace{6pt} 



\authorcontributions{Both authors contributed equally to all aspects of the manuscript.  All authors have read and agreed to the published version of the manuscript.}

\funding{A.K. was supported by NSF grants  DMR-2037654. T.G. acknowledges funding from the Institute of Science and Technology (IST) Austria, and from the European Union's Horizon 2020 research and innovation program under the Marie Sk\l{}odowska-Curie Grant Agreement No. 754411.}

\acknowledgments{We are indebted to Boris Shklovskii for introducing us to the problem, and Alexander Gorsky and Peter Koroteev for introducing us to the Picard-Fuchs methods. A very special thanks goes to Michael Janas for several years of excellent collaboration on these topics. TG thanks Michael Kreshchuk for introduction to the exact WKB method and great collaboration on related projects. Figures \ref{fig:Cycles11} and \ref{fig:Spectrum} are reproduced from Ref. \cite{JETP} with friendly permission by the Russian Academy of Sciences. Figures \ref{fig:TorusConstruction}, \ref{fig:Spectrum}, \ref{fig:CyclesNonHermitian}, \ref{fig:Quantization}, and \ref{fig:Bandwidth} are reproduced from Ref. \cite{JoPA} with friendly permission by IOP Publishing. \copyright IOP Publishing. All rights reserved.}

\conflictsofinterest{The authors declare no conflict of interest.}

\appendixtitles{yes} 
%

\reftitle{References}


\externalbibliography{yes}
\bibliography{IonChannelsReferences}




\end{document}